\documentclass[aps,prd,showpacs,twocolumn,groupedaddress]{revtex4}
\usepackage{graphicx}
\usepackage{amsfonts}
\usepackage{amssymb}
\usepackage{amsmath}

\usepackage{ulem}
\usepackage{color}

\newcommand{\Slash}[1]{{\ooalign{\hfil/\hfil\crcr$#1$}}}

\begin{document}

\title{Gauge-Invariant Formalism with a Dirac-mode Expansion \\
for Confinement and Chiral Symmetry Breaking}

\author{Shinya~Gongyo}
  \email{gongyo@ruby.scphys.kyoto-u.ac.jp}
  \affiliation{Department of Physics, Graduate School of Science,
  Kyoto University, \\
  Kitashirakawa-oiwake, Sakyo, Kyoto 606-8502, Japan}
\author{Takumi~Iritani}
  \email{iritani@ruby.scphys.kyoto-u.ac.jp}
  \affiliation{Department of Physics, Graduate School of Science,
  Kyoto University, \\
  Kitashirakawa-oiwake, Sakyo, Kyoto 606-8502, Japan}
\author{Hideo~Suganuma}
  \email{suganuma@ruby.scphys.kyoto-u.ac.jp}
  \affiliation{Department of Physics, Graduate School of Science,
  Kyoto University, \\
  Kitashirakawa-oiwake, Sakyo, Kyoto 606-8502, Japan}

\date{\today}
\begin{abstract}
Using the eigen-mode of the QCD Dirac operator 
$\Slash D=\gamma^\mu D^\mu$, we develop a manifestly 
gauge-covariant expansion and projection 
of the QCD operators such as the Wilson loop and the Polyakov loop. 
With this method, we perform a direct analysis of 
the correlation between confinement and chiral symmetry breaking 
in lattice QCD Monte Carlo calculation on $6^4$ at $\beta$=5.6. 
Even after removing the low-lying Dirac modes, 
which are responsible to chiral symmetry breaking, 
we find that the Wilson loop obeys the area law, 
and the slope parameter corresponding to the string tension or 
the confinement force is almost unchanged. 
We find also that the Polyakov loop remains to be almost zero 
even without the low-lying Dirac modes, 
which indicates the $Z_3$-unbroken confinement phase.
These results indicate that one-to-one correspondence does not 
hold for between confinement and chiral symmetry breaking 
in QCD. 
\end{abstract}
\maketitle

\section{Introduction}

Nowadays, quantum chromodynamics (QCD) has been established 
as the fundamental gauge theory of the strong interaction. 
However, nonperturbative properties of low-energy QCD 
such as color confinement and chiral symmetry breaking \cite{NJL61}
are not yet well understood, which gives 
one of the most difficult problems in theoretical physics. 
The nonperturbative QCD has been studied 
in lattice QCD \cite{W74,KS75,C7980,R05,Cr11} 
and various analytical frameworks 
\cite{N74,tH81,BC80,C82,S94,SST95}.

In particular, it is rather interesting and important 
to examine the correlation between confinement 
and chiral symmetry breaking 
\cite{SST95,M95,W95,HFGHO08,BBGH08,SWL08,C11,CPB11}, 
since the direct relation is not yet shown between them in QCD. 
The strong correlation between them has been suggested 
by the almost simultaneous phase transitions 
of deconfinement and chiral restoration in lattice QCD 
both at finite temperature \cite{R05,K02} 
and in a small-volume box \cite{R05}.

The close relation between confinement and chiral symmetry breaking 
has been also suggested in terms of the monopole 
degrees of freedom \cite{SST95,M95,W95}.
Here, the monopole topologically appears in QCD 
by taking the maximally Abelian (MA) gauge 
\cite{KSW87,SNW94,STSM95,KMS95,AS99}.
For example, by removing the monopoles in the MA gauge, 
confinement and chiral symmetry breaking are 
simultaneously lost in lattice QCD \cite{M95,W95}.
(The instantons also disappear without monopoles \cite{STSM95}.) 
This indicates an important role of the monopole 
to both confinement and chiral symmetry breaking, 
and these two nonperturbative QCD phenomena seem 
to be related via the monopole.
However, as a possibility, removing the monopoles 
may be ``too fatal'' for most nonperturbative properties. 
If this is the case, nonperturbative QCD 
phenomena are simultaneously lost by their cut. 

In fact, there remains an important question: 
{\it if only the relevant ingredient of 
chiral symmetry breaking is carefully removed, 
how will be confinement in QCD?}
In this paper, considering this question, 
we perform a direct investigation between 
color confinement and chiral symmetry breaking in lattice QCD, 
using the Dirac-mode expansion 
in a gauge-invariant manner \cite{SGIY11}.

The organization of this paper is as follows. 
In Sec.II, we introduce the gauge-invariant formalism 
with the Dirac-mode expansion. 
In Sec.III, we present the operator formalism in lattice QCD.
In Sec.IV, we formulate the Dirac-mode expansion and projection.
In Sec.V, we show the lattice results on the 
analysis of confinement in terms of the Dirac modes in QCD.
Section VI is devoted to summary and discussions.

\section{Gauge-Invariant Formalism with Dirac-mode Expansion}

We newly develop a manifestly gauge-covariant expansion 
of the QCD operator such as the Wilson loop, 
using the eigen-mode of the QCD Dirac operator 
$\Slash D=\gamma^\mu D^\mu$, and investigate 
the relation between confinement and chiral symmetry breaking. 

\subsection{Gauge Covariant Expansion in QCD 
instead of Fourier Expansion}

In the previous studies \cite{YS0809,YS10}, 
we investigated the relevant gluon-momentum region 
for confinement in lattice QCD, 
and found that the string tension $\sigma$, 
i.e., the confining force, is almost unchanged 
even after removing the high-momentum gluon component above 1.5GeV 
in the Landau gauge. 
In fact, the confinement property originates 
from the low-momentum gluon component below 1.5GeV, 
which is the upper limit to contribute to $\sigma$. 

The previous study on the relevant gluonic modes 
was based on the Fourier expansion, i.e., 
the eigen-mode expansion of the momentum operator $p^\mu$. 
Because of the commutable nature of $[p^\mu, p^\nu]=0$, 
all the momentum $p^\mu$ can be simultaneously diagonalized, 
which is one of the strong merits of the Fourier expansion. 
Also it keeps Lorentz covariance. 

However, the Fourier expansion does {\it not} 
keep gauge invariance in gauge theories.
Therefore, for the use of the Fourier expansion in QCD, 
one has to select a suitable gauge such as 
the Landau gauge \cite{YS0809,YS10,IS0911}, 
where the gauge-field fluctuation is strongly suppressed 
in Euclidean QCD.

As a next challenge, we consider a gauge-invariant method, 
using a gauge-covariant expansion in QCD instead of the 
Fourier expansion. 
In fact, we consider a generalization of the Fourier expansion or 
an alternative expansion with keeping the gauge symmetry. 

A straight generalization is to use the covariant derivative operator 
$D^\mu$ instead of the derivative operator $\partial^\mu$. 
However, due to the non-commutable nature of $[D^\mu, D^\nu] \ne 0$, 
one cannot diagonalize all the covariant derivative 
$D^\mu$ ($\mu=1,2,3,4$) simultaneously, 
but only one of them can be diagonalized. 
For example, the eigen-mode expansion of 
$D_4$ keeps gauge covariance and is rather interesting, 
but this type of expansion inevitably breaks the Lorentz covariance.
Then, we consider the eigen-mode expansion of 
the Dirac operator $\Slash D = \gamma^\mu D^\mu$ 
or $D^2 = D^\mu D^\mu$ \cite{BI05}, since such an expansion keeps both 
gauge symmetry and Lorentz covariance.

In particular, the Dirac-mode expansion is rather interesting because 
the Dirac operator $\Slash D$ directly connects with 
chiral symmetry breaking via the Banks-Casher relation \cite{BC80} 
and its zero modes are related to the topological charge 
via the Atiyah-Singer index theorem \cite{AS68}. 
Here, we mainly consider the manifestly gauge-invariant new method 
using the Dirac-mode expansion. 
Thus, the Dirac-mode expansion has some important merits:
\begin{itemize}
\item
The Dirac-mode expansion method manifestly keeps both 
gauge and Lorentz invariance.
\item
Each QCD phenomenon can be directly investigated in terms of 
chiral symmetry breaking.   
\end{itemize}

\subsection{Eigen-mode of Dirac Operator in Lattice QCD}

Now, we consider 
the Dirac operator and its eigen-modes 
in lattice QCD formalism with spacing $a$ in the Euclidean metric.
On the lattice, each site is labeled by 
$x=(x_1,x_2,x_3,x_4)$ with $x_\mu$ being an integer. 
In lattice QCD, the gauge field is described by 
the link-variable $U_\mu(x)=e^{iagA_\mu(x)}\in {\rm SU}(N_c)$, 
where $g$ is the QCD gauge coupling and 
$A_\mu(x)\in {\rm su}(N_c)$ corresponds to the gluon field. 

In lattice QCD, 
the Dirac operator 
$\Slash D = \gamma_\mu D_\mu$ is expressed with $U_\mu(x)$ as
\begin{eqnarray}
      \Slash{D}_{x,y} 
      = \frac{1}{2a} \sum_{\mu=1}^4 \gamma_\mu 
\left[ U_\mu(x) \delta_{x+\hat{\mu},y}
        - U_{-\mu}(x) \delta_{x-\hat{\mu},y} \right],
\end{eqnarray}
where the convenient notation 
$U_{-\mu}(x)\equiv U^\dagger_\mu(x-\hat \mu)$ is used.
Here, $\hat \mu$ denotes the unit vector 
on the lattice in $\mu$-direction \cite{R05}.

In this paper, we adopt the hermite definition of 
the $\gamma$-matrix, $\gamma_\mu^\dagger=\gamma_\mu$.
Thus, $\Slash D$ is anti-hermite and 
satisfies 
\begin{eqnarray}
\Slash D_{y,x}^\dagger=-\Slash D_{x,y}.
\end{eqnarray}
The normalized eigen-state $|n \rangle$ 
of the Dirac operator $\Slash D$ is introduced as 
\begin{eqnarray}
\Slash D |n\rangle =i \lambda_n |n \rangle
\end{eqnarray}
with $\lambda_n \in {\bf R}$.
Because of $\{\gamma_5,\Slash D\}=0$, the state 
$\gamma_5 |n\rangle$ is also an eigen-state of $\Slash D$ with the 
eigenvalue $-i\lambda_n$. 
The Dirac eigenfunction 
\begin{eqnarray}
\psi_n(x)\equiv\langle x|n \rangle
\end{eqnarray} 
obeys $\Slash D \psi_n(x)=i\lambda_n \psi_n(x)$, 
and its explicit form of the eigenvalue equation in lattice QCD is 
\begin{eqnarray}
&\frac{1}{2a}& \sum_{\mu=1}^4 \gamma_\mu
[U_\mu(x)\psi_n(x+\hat \mu)-U_{-\mu}(x)\psi_n(x-\hat \mu)]\nonumber \\
&=&i\lambda_n \psi_n(x).
\label{eq:eigen}
\end{eqnarray}
The Dirac eigenfunction $\psi_n(x)$ can be 
numerically obtained in lattice QCD, besides a phase factor. 

According to 
$U_\mu(x) \rightarrow V(x) U_\mu(x) V^\dagger (x+\hat\mu)$, 
the gauge transformation of $\psi_n(x)$ is found to be 
\begin{eqnarray}
\psi_n(x)\rightarrow V(x) \psi_n(x),
\label{eq:GTprop}
\end{eqnarray}
which is the same as that of the quark field.
To be strict, for the Dirac eigenfunction, 
there appears an irrelevant $n$-dependent global phase factor 
$e^{i\varphi_n[V]}$, 
according to the arbitrariness of the definition of $\psi_n(x)$.

It is notable that the quark condensate 
$\langle\bar qq \rangle$, the order parameter of 
chiral symmetry breaking, is given by 
the zero-eigenvalue density $\rho(0)$ 
of the Dirac operator, 
via the Banks-Casher relation \cite{BC80}, 
\begin{eqnarray}
\langle \bar qq \rangle=-\lim_{m \to 0} \lim_{V \to \infty} 
\pi\rho(0).
\end{eqnarray}
Here, the spectral density of the Dirac operator is defined by 
\begin{eqnarray}
\rho(\lambda)\equiv 
\frac1V\sum_{n}\langle \delta(\lambda-\lambda_n)\rangle,
\end{eqnarray} 
with the four-dimensional volume $V$.
Also, the zero-mode number asymmetry of 
the Dirac operator $\Slash D$ is equal to 
the topological charge (the instanton number)
$Q \equiv \frac{g^2}{16\pi^2}\int d^4x 
\ {\rm Tr} \ (G_{\mu\nu} \tilde G_{\mu\nu})$, 
which is known as the Atiyah-Singer index theorem, 
Index($\Slash D$)=$Q$ \cite{AS68}.


In calculating the eigenvalue of the Dirac operator $\Slash D$,
we use the Kogut-Susskind (KS) formalism \cite{KS75,R05}, 
which is often used to remove the redundant doublers of lattice fermions.
Here, the use of the KS formalism is just the practical reason 
to reduce the calculation of the Dirac eigenvalues. 
In fact, the result of the Dirac-mode projection, 
which will be shown in Sec.IV, is unchanged, 
when the Dirac operator is directly diagonalized.

In the KS method, using
$
T(x)\equiv \gamma_{1}^{x_1}\gamma_{2}^{x_2}
\gamma_{3}^{x_3}\gamma_{4}^{x_4}
$
with $\gamma_\mu^{-k}\equiv (\gamma_\mu^{-1})^k$ ($k=1,2, ...$),
all the gamma matrices $\gamma_\mu$ are diagonalized as 
$
T^\dagger(x) \gamma_\mu T(x \pm \hat \mu)=\eta_\mu(x)1 
$
with the staggered phase $\eta_\mu(x)$ defined by
\begin{eqnarray}
\eta_1(x)\equiv 1, \quad 
\eta_\mu(x)=(-1)^{x_1+\cdots+x_{\mu-1}} \ (\mu \ge 2). 
\end{eqnarray}
For $\chi_n(x) \equiv T^{\dagger}(x) \psi_n(x)$, 
the Dirac eigenvalue equation has no spinor index, 
and the spinor degrees of freedom can be dropped off, 
which reduces the lattice-fermion species from 16 to 4 \cite{R05}. 
In the KS method, 
the Dirac operator $\gamma_\mu D_\mu$ is 
replaced by the KS Dirac operator $\eta_\mu D_\mu$, 
\begin{eqnarray}
(\eta_\mu D_\mu)_{x,y} = 
\frac{1}{2a} \sum_{\mu=1}^4 
\eta_\mu(x)[U_\mu(x) \delta_{x+\hat \mu,y} 
-U_{-\mu}(x) \delta_{x-\hat \mu,y}],
\end{eqnarray}
and the spinless eigenfunction $\chi_n(x)$ satisfies 
\begin{eqnarray}
\frac{1}{2a}\sum_{\mu=1}^4 
\eta_\mu(x)[U_\mu(x) \chi_n(x+\hat \mu)-U_{-\mu}(x)
\chi_n(x-\hat \mu)] \nonumber \\
=i\lambda_n\chi_n(x).
\label{eq:KSeigen}
\end{eqnarray}
In the KS formalism, 
the chiral partner $\gamma_5 \psi_n(x)$ 
reduces into $\eta_5(x)\chi_n(x)=(-1)^{x_1+x_2+x_3+x_4}\chi_n(x)$, 
which is an eigenfunction of $\eta_\mu D_\mu$ 
with the eigenvalue $-i\lambda_n$.

Using the KS formalism \cite{KS75,R05}, 
the Dirac-mode number $L^4 \times N_c \times$ 4 
is reduced to be $L^4 \times N_c$ on the $L^4$ lattice. 
The actual number of the independent Dirac eigenvalue 
$\lambda_n$ is about $L^4 \times N_c/2$, 
due to the chiral property of the Dirac operator, i.e., 
pairwise appearance of $\pm \lambda_n$.

\section{Operator Formalism in Lattice QCD}

To keep the gauge symmetry, careful treatments are necessary, 
since naive approximations may break the gauge symmetry. 
Here, we take the ``operator formalism'' \cite{SGIY11}, 
as explained below.

We define the link-variable operator $\hat U_{\pm \mu}$ 
by the matrix element of 
\begin{eqnarray}
\langle x |\hat U_{\pm \mu}|y\rangle
=U_{\pm \mu}(x)\delta_{x \pm \hat \mu,y}.
\end{eqnarray}
Note that $\hat U_{\mu}$ and $\hat U_{-\mu}$ are 
Hermitian conjugate as the operator in the Hilbert space 
in the sense that
\begin{eqnarray}
\langle y |\hat U_\mu^\dagger|x\rangle 
&=&U_{\mu}^\dagger(y)\delta_{y+\hat \mu, x}
=U_{\mu}^\dagger(x-\hat \mu)\delta_{x-\hat \mu, y}\nonumber \\
&=&U_{-\mu}(x)\delta_{x-\hat \mu, y}
=\langle x |\hat U_{-\mu}|y \rangle.
\end{eqnarray}
In the operator formalism, 
Eq.(\ref{eq:eigen}) for the Dirac eigen-state 
is simply expressed as
\begin{eqnarray}
\frac{1}{2a} \sum_{\mu=1}^4 \gamma_\mu
(\hat U_\mu -\hat U_{-\mu}) |n \rangle =i\lambda_n |n \rangle.
\end{eqnarray}

In the KS method, where the spinor index is dropped off, 
one identifies $\chi_n(x)=\langle x|n\rangle$, 
and then Eq.(\ref{eq:KSeigen}) 
for the KS Dirac eigen-state is expressed as 
\begin{eqnarray}
\frac{1}{2a} \sum_{\mu=1}^4 \hat \eta_\mu 
(\hat U_\mu -\hat U_{-\mu}) |n \rangle =i\lambda_n |n \rangle,
\label{eq:KSop}
\end{eqnarray}
where 
$\hat \eta_\mu$ 
is defined by 
$
\langle x|\hat \eta_\mu|y \rangle=\eta_\mu(x)\delta_{x,y}.
$
Owing to $\eta_\mu(x \pm \hat \mu)=\eta_\mu(x)$, 
one finds $\hat \eta_\mu \hat U_{\pm \mu} 
=\hat U_{\pm \mu} \hat \eta_\mu$,
so that there is no ordering uncertainty in 
the KS Dirac operator in Eq.(\ref{eq:KSop}).
In the KS method, the chiral partner $\gamma_5|n \rangle$ 
corresponds to $\hat \eta_5|n \rangle$, where 
$\hat \eta_5$ is defined by the matrix element 
$\langle x|\hat \eta_5|y\rangle=\eta_5(x)\delta_{x,y}
=(-1)^{x_1+x_2+x_3+x_4}\delta_{x,y}$.
Due to $\eta_5(x \pm \hat \mu)=-\eta_5(x)$, we note 
$\hat \eta_5\hat U_{\pm \mu}=-\hat U_{\pm \mu} \hat \eta_5$.

In the following, we mainly use 
the ordinary Dirac operator $\gamma_\mu D_\mu$ 
and the spinor eigenfunction $\psi_n(x)=\langle x|n\rangle$. 
When the KS method is applied, one only has to 
use the identification of $\chi_n(x)=\langle x|n\rangle$ 
in the following arguments. 
The final results are the same between both calculations 
based on $\gamma_\mu D_\mu$ and $\eta_\mu D_\mu$.

The Wilson-loop operator $\hat W$ is defined as the product of 
$\hat U_\mu$ along a rectangular loop,
\begin{eqnarray}
\hat W \equiv \prod_{k=1}^N \hat U_{\mu_k}
=\hat U_{\mu_1}\hat U_{\mu_2} \cdots \hat U_{\mu_N}.
\end{eqnarray}
For arbitrary loops, one finds $\sum_{k=1}^N \hat \mu_k=0$.
We note that the functional trace of the Wilson-loop operator 
$\hat W$ is proportional to the ordinary vacuum expectation value 
$\langle W \rangle$ of the Wilson loop:
\begin{widetext}
\begin{eqnarray}
{\rm Tr} \ \hat W&=&{\rm tr}\sum_x \langle x |\hat W|x \rangle
={\rm tr}\sum_x \langle x| \hat U_{\mu_1}\hat U_{\mu_2} 
\cdots \hat U_{\mu_N}|x\rangle \nonumber\\
&=& {\rm tr} \sum_{x_1, x_2, \cdots, x_N }
\langle x_1| \hat U_{\mu_1}|x_2 \rangle
\langle x_2| \hat U_{\mu_2}|x_3 \rangle
\langle x_3| \hat U_{\mu_3}|x_4 \rangle
\cdots \langle x_N|\hat U_{\mu_N}|x_1\rangle \nonumber\\
&=&{\rm tr} \sum_x 
\langle x| \hat U_{\mu_1}|x+\hat \mu_1 \rangle
\langle x+\hat \mu_1| \hat U_{\mu_2}|x+\sum_{k=1}^2\hat \mu_k \rangle
\cdots \langle x+\sum_{k=1}^{N-1}\hat \mu_k|\hat U_{\mu_N}|x\rangle \nonumber\\
&=&\sum_x {\rm tr}\{ U_{\mu_1}(x) U_{\mu_2}(x+\hat \mu_1)
U_{\mu_3}(x+\sum_{k=1}^2 \hat \mu_k)
\cdots U_{\mu_N}(x+\sum_{k=1}^{N-1} \hat \mu_k)\} \nonumber\\
&=&\langle W \rangle \cdot {\rm Tr}\ 1.
\label{eq:TrWLO}
\end{eqnarray}
\end{widetext}
Here, ``Tr'' denotes the functional trace, 
and ``tr'' the trace over SU(3) color index.

The Dirac-mode matrix element of the link-variable operator 
$\hat U_{\mu}$ can be expressed with $\psi_n(x)$:
\begin{eqnarray}
\langle m|\hat U|n \rangle&=&\sum_x\langle m|x \rangle 
\langle x|\hat U_{\mu}|x
+\hat \mu \rangle \langle x+\hat \mu|n\rangle \nonumber \\
&=&\sum_x \psi_m^\dagger(x) U_\mu(x)\psi_n(x+\hat \mu).
\end{eqnarray}
Although the total number of the matrix element is very huge, 
the matrix element is calculable and gauge invariant, 
apart from an irrelevant phase factor.
Using the gauge transformation (\ref{eq:GTprop}), we find 
the gauge transformation of the matrix element as 
\begin{widetext}
\begin{eqnarray}
\langle m|\hat U_\mu|n \rangle
&=&\sum_x \psi^\dagger_m(x)U_\mu(x)\psi_n(x+\hat\mu) \nonumber\\
&\rightarrow&
\sum_x\psi^\dagger_m(x)V^\dagger(x)\cdot V(x)U_\mu(x)V^\dagger(x+\hat \mu)
\cdot V(x+\hat \mu)\psi_n(x+\hat \mu) \nonumber\\
&=&\sum_x\psi_m^\dagger(x)U_\mu(x)\psi_n(x+\hat \mu)
=\langle m|\hat U_\mu|n\rangle.
\end{eqnarray}
\end{widetext}
To be strict, there appears an $n$-dependent global phase factor, 
corresponding to the arbitrariness of the phase in the basis 
$|n \rangle$. However, this phase factor cancels as 
$e^{-i\varphi_n} e^{i\varphi_n}=1$ 
between $|n \rangle$ and $\langle n |$, and does not appear 
for QCD physical quantities including 
the Wilson loop and the Polyakov loop.

In the practical lattice-QCD calculation, 
we adopt the KS formalism to reduce the computational complexity, 
as mentioned in Sec II-B. 
In the KS method, instead of $\psi_n(x)$, 
we use the spinless eigenfunction $\chi_n(x)$ of 
the KS Dirac operator $\eta_\mu D_\mu$, 
with the identification of $\chi_n(x)=\langle x|n\rangle$,
and the KS-reduced matrix element of $\hat U_\mu$ is expressed as 
\begin{eqnarray}
\langle m|\hat U|n \rangle&=&\sum_x\langle m|x \rangle 
\langle x|\hat U_{\mu}|x
+\hat \mu \rangle \langle x+\hat \mu|n\rangle \nonumber \\
&=&\sum_x \chi_m^\dagger(x) U_\mu(x)\chi_n(x+\hat \mu).
\end{eqnarray}
In the arguments in the next section, 
the same results are obtained between the calculations 
based on the original Dirac operator $\gamma_\mu D_\mu$ 
and the KS Dirac operator $\eta_\mu D_\mu$.

\section{Dirac-mode Expansion and Projection}

\subsection{General Definition of Dirac-mode Expansion and Projection}

From the completeness of the Dirac-mode basis, 
$\sum_n|n\rangle \langle n|=1$, we get 
\begin{eqnarray}
\hat O=\sum_m\sum_n |m \rangle \langle m|\hat O|n \rangle \langle n|
\end{eqnarray} 
for arbitrary operators.
Based on this relation, the Dirac-mode expansion and projection 
can be defined \cite{SGIY11}. 
We define the projection operator $\hat P$ 
which restricts the Dirac-mode space, 
\begin{eqnarray}
\hat P\equiv \sum_{n \in A}|n\rangle \langle n|,
\end{eqnarray} 
where $A$ denotes arbitrary set of Dirac modes. 
In $\hat P$, the arbitrary phase cancels 
between $|n\rangle$ and $\langle n|$. 
One finds $\hat P^2=\hat P$ and $\hat P^\dagger =\hat P$.
The typical projections are 
IR-cut and UV-cut of the Dirac modes:
\begin{eqnarray}
\hat P_{\rm \ IR} \equiv 
\sum_{|\lambda_n| \ge \Lambda_{\rm IR}}|n \rangle \langle n|,
\quad 
\hat P_{\rm \ UV} \equiv 
\sum_{|\lambda_n| \le \Lambda_{\rm UV}}|n \rangle \langle n|.
\end{eqnarray} 

Using the projection operator $\hat P$, we define 
the Dirac-mode projected link-variable operator, 
\begin{eqnarray}
\hat U^P_\mu \equiv \hat P \hat U_\mu \hat P
=\sum_{m \in A}\sum_{n \in A} 
|m\rangle \langle m|\hat U_\mu|n\rangle \langle n|.
\end{eqnarray}
During this projection, there appears some non-locality in general, 
but it would not be important for the argument of 
large-distance properties such as confinement. 
%

Each lattice QCD configuration is characterized by the 
set of the link-variable $\{U_\mu(s)\}$, or equivalently, 
the link-variable operator $\{\hat U_\mu\}$, and then 
the Dirac-mode projection is described by the 
replacement of $\{\hat U_\mu\}$ by $\{\hat U^P_\mu\}$.
In fact, the Dirac-mode projection 
of QCD physical quantities $\langle O[U_\mu(s)] \rangle$ or 
${\rm Tr} \hat O[\hat U_\mu]$ can be defined 
by the replacement of 
\begin{eqnarray}
{\rm Tr} \hat O[\hat U_\mu] \rightarrow {\rm Tr} \hat O[\hat U^P_\mu].
\end{eqnarray}
Also in full QCD, after the integration over the quark degrees of freedom, 
all the QCD physical quantities can be written 
by $\langle O[U_\mu(s)] \rangle$ or ${\rm Tr} \hat O[\hat U_\mu]$, 
so that the Dirac-mode projection can be applied in the same way.

\subsection{Dirac-mode expansion and projection of the Wilson loop}

In this subsection, we consider the Dirac-mode 
expansion and projection of the Wilson-loop 
$\langle W (R,T) \rangle \propto {\rm Tr} \hat W(R,T)$ 
corresponding to the $R \times T$ rectangular loop.

For the ordinary Wilson loop $\langle W (R,T) \rangle$, 
its area law indicates the confinement phase of 
the QCD vacuum and the linear arising potential between static 
quark and antiquark in the infrared region \cite{R05}.

From the Wilson-loop operator 
$
\hat W \equiv \prod_{k=1}^N\hat U_{\mu_k}, 
$
we get the Dirac-mode expansion of the Wilson loop as 
\begin{widetext}
\begin{eqnarray}
{\rm Tr} \hat W = {\rm Tr} \prod_{k=1}^N \hat U_{\mu_k}
={\rm Tr} ( \hat U_{\mu_1}\hat U_{\mu_2}\cdots \hat U_{\mu_N} )
={\rm tr} \sum_{n_1, n_2, \cdots, n_{N}} 
\langle n_1| \hat U_{\mu_1}
|n_2 \rangle \langle n_2| \hat U_{\mu_2} |n_3 \rangle \cdots
\langle n_N| \hat U_{\mu_N}|n_{1} \rangle.
\end{eqnarray}
\end{widetext}
Based on this expression, we investigate the role of 
specific Dirac modes to the area law of the Wilson loop. 
In fact, if some Dirac modes are essential to reproduce 
the area law of the Wilson loop or the confinement property, 
the removal of the coupling to these modes leads to 
a significant change on the area law.

In this way, we try to answer the question of 
{\it ``Are there any relevant Dirac modes responsible 
to the area law of the Wilson loop?''} 

To this end, we define the Dirac-mode projected 
Wilson-loop operator,
\begin{widetext}
\begin{eqnarray}
\hat W^P &\equiv& \prod_{k=1}^N \hat U^P_{\mu_k}
=\hat U^P_{\mu_1}\hat U^P_{\mu_2}\cdots \hat U^P_{\mu_N} 
=\hat P \hat U_{\mu_1} \hat P \hat U_{\mu_2} \hat P 
\cdots \hat P \hat U_{\mu_N} \hat P \nonumber\\
&=&\sum_{n_1, n_2, \cdots, n_{N+1} \in A} 
|n_1 \rangle \langle n_1| \hat U_{\mu_1}
|n_2 \rangle \langle n_2| \hat U_{\mu_2} |n_3 \rangle \cdots
\langle n_N| \hat U_{\mu_N}
|n_{N+1} \rangle \langle n_{N+1}|.
\end{eqnarray}
Then, we obtain the functional trace of 
the Dirac-mode projected Wilson-loop operator, 
\begin{eqnarray}
{\rm Tr} \ \hat W^P &=& {\rm Tr} \ \prod_{k=1}^N \hat U^P_{\mu_k} 
={\rm Tr} \ \hat U^P_{\mu_1}\hat U^P_{\mu_2}\cdots \hat U^P_{\mu_N} 
={\rm Tr} \ \hat P \hat U_{\mu_1} \hat P \hat U_{\mu_2} \hat P 
\cdots \hat P \hat U_{\mu_N} \hat P \nonumber\\
&=&{\rm tr} \sum_{n_1, n_2, \cdots, n_N \in A} 
\langle n_1| \hat U_{\mu_1} |n_2 \rangle 
\langle n_2| \hat U_{\mu_2} |n_3 \rangle \cdots
\langle n_N| \hat U_{\mu_N}|n_{1} \rangle,
\end{eqnarray}
\end{widetext}
which is manifestly gauge invariant. 
Here, the arbitrary phase factor 
cancels between $|n_k \rangle$ and $\langle n_k|$. 
Its gauge invariance is also numerically checked 
in the lattice QCD Monte Carlo calculation.

The original Wilson-loop operator $\hat W(R,T)$ couples to 
all the Dirac modes, and 
${\rm Tr} \ \hat W(R,T)$ obeys the area law, 
\begin{eqnarray}
{\rm Tr} \ \hat W(R,T) \propto \langle W (R,T) \rangle \propto e^{-\sigma RT},
\end{eqnarray}
for large $R$ and $T$. Here, the slope parameter $\sigma$ 
corresponds to the string tension or the confinement force.
For the restriction of the Dirac-mode space to be $A$, 
we investigate the Dirac-mode projected Wilson-loop operator 
$\hat W^P(R,T)$, which couples to the restricted Dirac modes.
If the removed Dirac modes are essential for the confinement property or 
the area-law behavior of the Wilson loop, a large change is expected 
on the behavior of ${\rm Tr} \ \hat W^P(R,T)$. 
If not, no significant change is expected on the behavior 
of ${\rm Tr} \ \hat W^P(R,T)$.
In fact, one can investigate the role of the removed Dirac modes 
to confinement by checking the area-law behavior of 
${\rm Tr} \ \hat W^P(R,T)$ and the slope parameter $\sigma^P$, 
which is formally written as 
\begin{eqnarray}
\sigma^P \equiv -\lim_{R,T \rightarrow \infty } 
\frac1{RT}{\rm ln} \{ {\rm Tr} \ \hat W^P(R,T)\}.
\label{eq:slopedef}
\end{eqnarray}

\subsection{Corresponding Dirac-mode projected inter-quark potential}

For the estimation of 
the slope parameter $\sigma^P$ from ${\rm Tr} \ \hat W^P(R,T)$, 
we define the corresponding Dirac-mode projected inter-quark potential, 
\begin{eqnarray}
V^P(R)\equiv -\lim_{T \to \infty} \frac{1}{T}
{\rm ln} \{{\rm Tr} \ \hat W^P(R,T)\},
\end{eqnarray}
which is also manifestly gauge-invariant.
To be strict, due to the non-locality 
appearing in the Dirac-mode projection, 
$V^P(R)$ does not have a definite meaning of 
the static potential. 
However, it is still useful to obtain $\sigma^P$ 
in Eq.(\ref{eq:slopedef}) from ${\rm Tr} \hat W^P(R,T)$. 
In fact, $\sigma^P$ is obtained from the infrared slope of $V^P(R)$.
Note also that, in the unprojected case of $\hat P=1$, 
the ordinary inter-quark potential is obtained 
apart from an irrelevant constant,
\begin{eqnarray}
V(R)&=&-\lim_{T \to \infty} \frac{1}{T}
{\rm ln} \{{\rm Tr} \ \hat W(R,T)\} \nonumber \\
&=& -\lim_{T \to \infty} \frac{1}{T}
{\rm ln} \langle W(R,T)\rangle + {\rm irrelevant} \ {\rm const.},~~~~~~
\label{eq:potdef}
\end{eqnarray}
because of ${\rm Tr} \ \hat W=\langle W \rangle 
\cdot {\rm Tr}\ 1$, as was derived in Eq.(\ref{eq:TrWLO}).

\section{Analysis of Confinement in terms of Dirac Modes in QCD}

We consider various projection space $A$ in the Dirac-mode space, 
e.g., IR-cut or UV-cut of Dirac modes. 
With this Dirac-mode expansion and projection formalism, 
we calculate the Dirac-mode projected Wilson loop ${\rm Tr} W^P(R,T)$ 
in a gauge-invariant manner. 
In particular, using IR-cut of the Dirac modes, 
we directly investigate the relation between chiral symmetry breaking 
and confinement as the area-law behavior of the Wilson loop, 
since the low-lying Dirac modes are 
responsible to chiral symmetry breaking.

As a technical difficulty of this formalism, we have to deal with 
huge dimensional matrices and their products.  
Actually, the total matrix dimension of 
$\langle m|\hat U_\mu|n\rangle$ is (Dirac-mode number)$^2$. 
On the $L^4$ lattice, the Dirac-mode number is 
$L^4 \times N_c \times$ 4, which can be reduced to be $L^4 \times N_c$, 
using the KS formalism \cite{KS75,R05}, as mentioned in Sec.II-B. 
The actual number of the independent Dirac eigenvalue 
$\lambda_n$ is about $L^4 \times N_c/2$, 
due to the chiral property of $\Slash D$, i.e., 
pairwise appearance of $\pm \lambda_n$.
Even for the projected operators, where the Dirac-mode space is 
restricted, the matrix is generally still huge. 
In addition, we have to deal with the product of 
the huge matrices $\langle m|\hat U_\mu|n \rangle$ 
in calculating the Wilson loop.
Then, at present, we use a small-size lattice 
in the numerical calculation.

In this paper, we perform the SU(3) lattice QCD Monte Carlo calculation 
with the standard plaquette action 
at $\beta=5.6$ on $6^4$ at the quenched level, 
using the pseudo-heat-bath algorithm. 
The ordinary periodic boundary condition is used for the link-variable. 
The gauge configurations are taken 
every 500 sweeps after 10,000 sweeps thermalization, 
and 20 gauge configurations are used for each analysis.
At $\beta=5.6$, the lattice spacing $a$ is estimated as 
$a\simeq 0.25{\rm fm}$, i.e., $a^{-1}\simeq 0.8{\rm GeV}$, 
which leads to the string tension $\sigma \simeq 0.89{\rm GeV/fm}$ 
in the inter-quark potential.
(This estimate is done also on a larger volume lattice.)
Then, the total volume is $V=(6a)^4\simeq(1.5{\rm fm})^4$, 
and the momentum cutoff is $\pi/a \simeq 2.5{\rm GeV}$.

On the $6^4$ lattice, the Dirac-mode number is 
$6^4 \times 3 \times 4=15,552$, which is reduced 
to be $6^4 \times 3=3,888$ using the KS formalism.
In fact, the KS Dirac operator $\eta_\mu D_\mu$ and 
the KS-reduced matrix element $\langle m|\hat U_\mu|n\rangle$
are expressed by $3,888 \times 3,888$ matrix. 
Considering the pairwise appearance of $\lambda_n$ and $-\lambda_n$, 
the actual number of the independent Dirac eigenvalue $\lambda_n$ 
is reduced to be around $6^4 \times 3/2=1,944$.

To diagonalize the KS Dirac operator $\eta_\mu D_\mu$, 
we use LAPACK \cite{LAPACK}.
For the statistical error on the lattice data, 
we adopt the jackknife error estimate \cite{R05}.

We show in Fig.1(a) the spectral density $\rho(\lambda)$ of 
the QCD Dirac operator $\Slash D$. 
The chiral property of $\Slash D$ leads to 
$\rho(-\lambda)=\rho(\lambda)$.
Figure~1(b) is the IR-cut Dirac spectral density 
\begin{eqnarray}
\rho_{\rm IR}(\lambda)\equiv 
\rho(\lambda)\theta(|\lambda|-\Lambda_{\rm IR})
\end{eqnarray}
with the IR-cutoff $\Lambda_{\rm IR}=0.5a^{-1}\simeq 0.4{\rm GeV}$.

\begin{figure}[h]
\begin{center}
\includegraphics[scale=0.5]{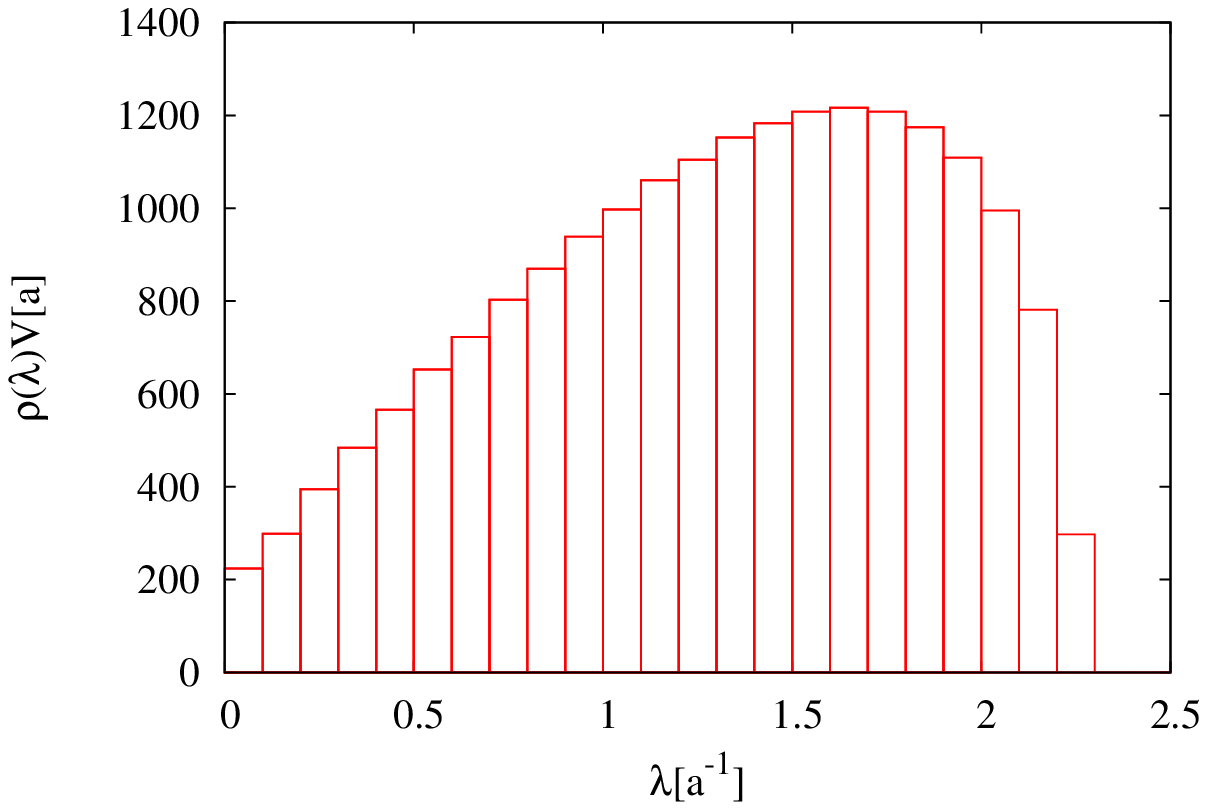}
\includegraphics[scale=0.5]{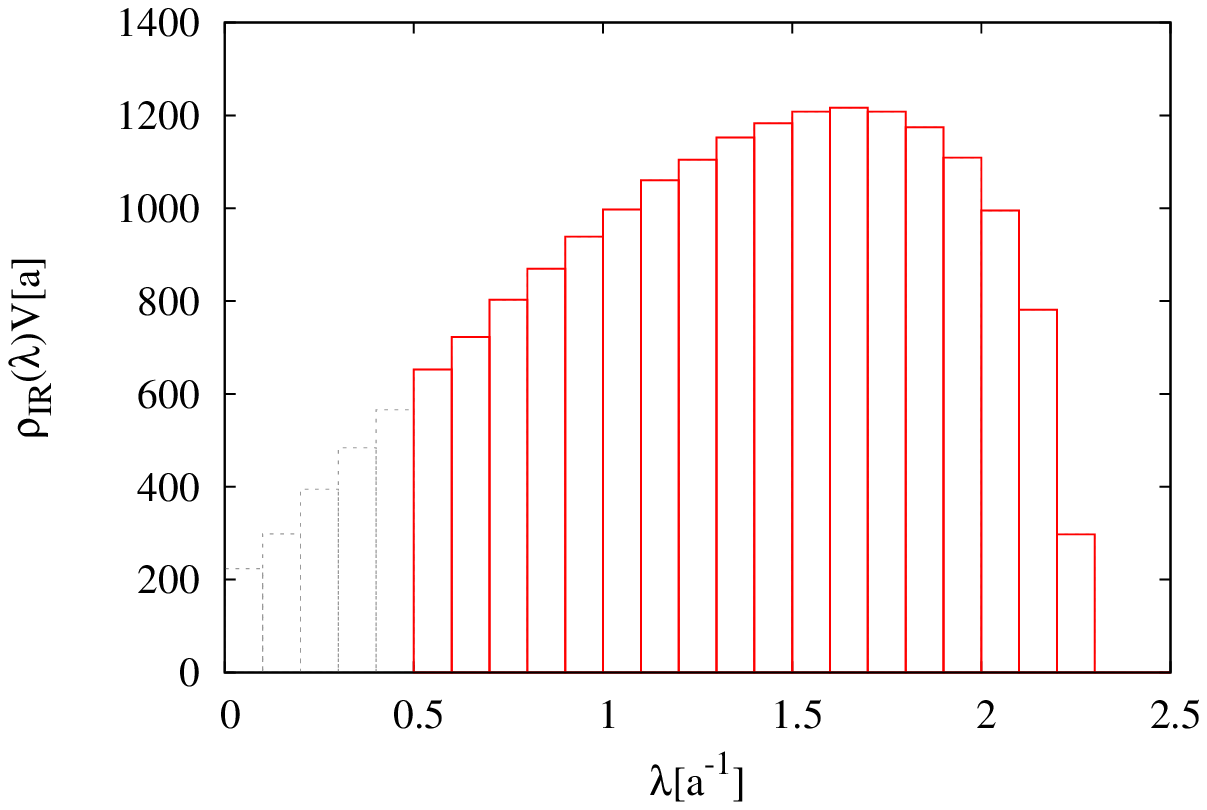}
\caption{
(a) The spectral density $\rho(\lambda)$ 
of the Dirac operator in lattice QCD at $\beta$=5.6 and $6^4$.
The negative-$\lambda$ region is omitted, because of 
$\rho(-\lambda)=\rho(\lambda)$. 
The volume $V$ is multiplied.
(b) The IR-cut Dirac spectral density 
$\rho_{\rm IR}(\lambda)\equiv 
\rho(\lambda)\theta(|\lambda|-\Lambda_{\rm IR})$ 
with the IR-cutoff $\Lambda_{\rm IR}=0.5a^{-1}\simeq 0.4{\rm GeV}$.
}
\end{center}
\end{figure}

Note that, using the eigenvalue $\lambda_n$, 
the quark condensate 
$\langle \bar qq\rangle$ is obtained as 
\begin{eqnarray}
\langle \bar qq\rangle
&=&-\frac{1}{V}{\rm Tr}\frac{1}{\Slash D+m}
=-\frac{1}{V}\sum_n\frac{1}{i\lambda_n+m} \nonumber \\
&=&-\frac{1}{V}\left(\sum_{\lambda_n>0} \frac{2m}{\lambda_n^2+m^2}
+\frac{\nu}{m}\right),
\end{eqnarray}
where $\nu$ is the total number of the zero mode of $\Slash D$.
Here, the non-zero eigenvalues appear as pairwise, 
which makes $\langle \bar qq \rangle$ real.
(In lattice QCD, one has to take account of the doubler contribution, 
which can be regarded as flavor at the quenched level.)
Then, in the presence of the IR cut $\Lambda_{\rm IR}$ 
for the Dirac eigen-mode, the quark condensate is obtained as 
\begin{eqnarray}
\langle \bar qq\rangle_{\Lambda_{\rm IR}}
=-\frac{1}{V}\sum_{\lambda_n \ge \Lambda_{\rm IR}} \frac{2m}{\lambda_n^2+m^2}.
\end{eqnarray}
We show in Fig.2 the lattice QCD result of 
the quark condensate $\langle \bar qq\rangle_{\Lambda_{\rm IR}}$ 
as the function of the current quark mass $m$ 
in the presence of IR cut $\Lambda_{\rm IR}$.

\begin{figure}[h]
\begin{center}
\includegraphics[scale=0.5]{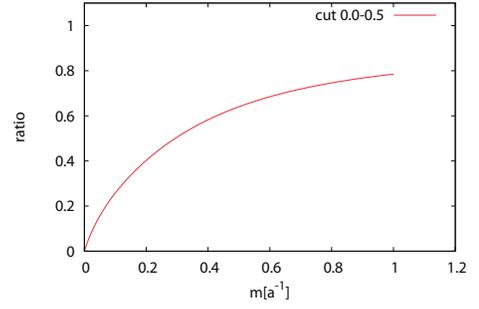}
\caption{
The lattice QCD result of 
the quark condensate $\langle \bar qq\rangle_{\Lambda_{\rm IR}}$ 
as the function of the current quark mass $m$
in the presence of IR cut $\Lambda_{\rm IR}=0.5 a^{-1}$.
The vertical axis is normalized by 
the original value of $\langle \bar qq\rangle$ without cut.
A large reduction is found as 
$\langle \bar qq\rangle_{\Lambda_{\rm IR}}/\langle \bar qq\rangle 
\simeq 0.02$
for $\Lambda_{\rm IR}=0.5a^{-1} \simeq 0.4{\rm GeV}$ 
around the physical region of $m \simeq 0.006a^{-1} \simeq 5{\rm MeV}$.
}
\end{center}
\end{figure}

By removing the low-lying Dirac modes, 
the chiral condensate $\langle \bar qq \rangle$ 
is largely reduced, reflecting the Banks-Casher relation. 
Actually, directly from lattice QCD calculation, 
we find a large reduction of the chiral condensate 
in the presence of the IR cut 
$\Lambda_{\rm IR}=0.5a^{-1}\simeq 0.4{\rm GeV}$,
\begin{eqnarray}
\frac{\langle \bar qq\rangle_{\Lambda_{\rm IR}}}
{\langle \bar qq\rangle} \simeq 0.02,
\end{eqnarray}
around the physical region of 
$m \simeq 0.006a^{-1} \simeq 5{\rm MeV}$ \cite{PDG}, as shown Fig.2.

Now, let us consider the removal of 
the coupling to the low-lying Dirac modes 
from the Wilson loop $\langle W(R,T)\rangle$. 
Figure~3 shows the Dirac-mode projected 
Wilson loop $\langle W^P(R,T) \rangle \equiv {\rm Tr} \hat W^P(R,T)$ 
after removing low-lying Dirac modes, 
which is obtained in lattice QCD with the IR-cut of 
$\rho_{\rm IR}(\lambda)\equiv 
\rho(\lambda)\theta(|\lambda|-\Lambda_{\rm IR})$ 
with the IR-cutoff $\Lambda_{\rm IR}=0.5a^{-1}$.
Even after removing the coupling to the low-lying Dirac modes, 
which are responsible to chiral symmetry breaking, 
the Dirac-mode projected Wilson loop is found to 
obey the area law as 
\begin{eqnarray}
\langle W^P(R,T)\rangle \propto e^{-\sigma^P RT},
\end{eqnarray}
and the slope parameter $\sigma^P$ 
corresponding to the string tension 
or the confinement force is almost unchanged as 
\begin{eqnarray}
\sigma^P \simeq \sigma.
\end{eqnarray}
In fact, the confinement property seems to be kept 
in the absence of the low-lying Dirac modes or 
the essence of chiral symmetry breaking \cite{SGIY11}. 
This result indicates that one-to-one correspondence does not 
hold for confinement and chiral symmetry breaking in QCD.

\begin{figure}[h]
\begin{center}
\includegraphics[scale=0.65]{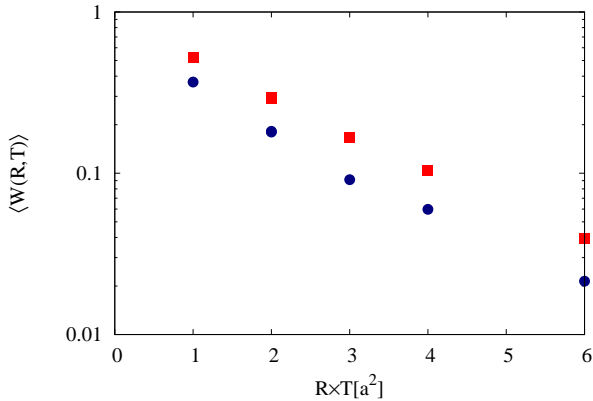}
\caption{
The lattice QCD result of the Dirac-mode projected Wilson loop 
$\langle W^P (R,T)\rangle \equiv {\rm Tr}\hat W^P(R,T)$ 
after removing low-lying Dirac modes, 
plotted against the area $R \times T$. 
The circle symbol denotes the Wilson loop 
obtained with the IR-cut of 
$\rho_{\rm IR}(\lambda)\equiv 
\rho(\lambda)\theta(|\lambda|-\Lambda_{\rm IR})$ 
with the IR-cutoff $\Lambda_{\rm IR}=0.5a^{-1}$.
The square symbol denotes the original Wilson loop 
$\langle W(R,T) \rangle$.
$\langle W^P (R,T)\rangle$ seems to obey the area law 
with the same slope parameter, $\sigma^P \simeq \sigma$. 
}
\end{center}
\end{figure}

Next, to estimate the slope parameter $\sigma^P$, 
we consider the potential $V^P(R)$ obtained 
from $\langle W^P(R,T)\rangle$. 
Figure~4 shows 
the ``effective mass'' of the inter-quark potential 
$
V_{\rm eff}(R,T) \equiv 
{\rm ln}[\langle W^P(R,T)\rangle/\langle W^P(R,T+1)\rangle]
$
after removing the low-lying Dirac modes, 
plotted against $T$ at each $R$.
One finds the ``plateau'' or the stability of 
the effective mass $V_{\rm eff}(R,T)$ against $T$, 
which means the dominance of the ground-state component.
\begin{figure}[h]
\begin{center}
\includegraphics[scale=0.5]{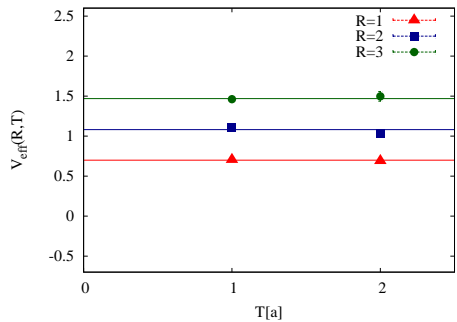}
\caption{
The effective-mass plot of the inter-quark potential 
$V_{\rm eff}(R,T) \equiv 
{\rm ln}[\langle W^P(R,T)\rangle/\langle W^P(R,T+1)\rangle]$ 
after removing the low-lying Dirac modes. 
$V_{\rm eff}(R,T)$ is plotted against $T$ for each $R$.
The horizontal line denotes the best-fit value in 
the exponential fit of Eq.(\ref{eq:fit}) at each $R$.
}
\end{center}
\end{figure}
Similarly in the standard procedure 
to obtain potentials in lattice QCD \cite{R05,TS02}, 
we determine the inter-quark potential $V^P(R)$ 
by the exponential fit of the Wilson loop 
\begin{eqnarray}
\langle W^P(R,T)\rangle=C e^{-V^P(R)T}
\label{eq:fit}
\end{eqnarray}
for $T=1,2,3$, which corresponds to 
the plateau region of $T=1,2$ in $V_{\rm eff}(R,T)$. 
%
Figure~5 shows the Dirac-mode projected 
inter-quark potential $V^P(R)$ 
after removing low-lying Dirac modes below 
the IR-cutoff $\Lambda_{\rm IR}=0.5a^{-1}$.
No significant change is observed 
on the inter-quark potential besides an irrelevant constant, 
that is, the slope parameter $\sigma^P$ is almost unchanged, 
even after removing the low-lying Dirac modes.

\begin{figure}[h]
\begin{center}
\includegraphics[scale=0.6]{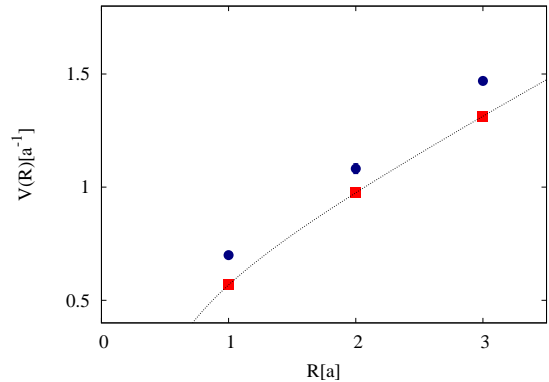}
\caption{
The lattice QCD result (circle) of the inter-quark potential $V^P(R)$ 
after removing low-lying Dirac modes 
below the IR-cutoff $\Lambda_{\rm IR}=0.5a^{-1}$.
The square symbol denotes the original inter-quark potential.
The potential is almost unchanged 
even after removing the low-lying Dirac modes,
apart from an irrelevant constant.
}
\end{center}
\end{figure}

On the potential argument, 
we comment on the non-locality stemming from the Dirac-mode projection,
which makes the link-variable extended and makes the potential meaning vague. 
This non-locality appears hyper-cubic symmetrically in the four-dimensional 
space-time, and its effect would be maximal for the IR-cut case.
As a whole, such a non-locality makes the potential flat, 
due to the spatial averaging. (As an extreme example, the ``potential'' 
between wall-like sources is completely flat.)
However, our obtained potential is almost the same as the original 
confining one, in spite of the possible flattening effect by the non-locality. 
Therefore, regardless of the non-locality, the confinement is kept 
after cutting off the low-lying Dirac modes. 
(Since no flattening effect is observed in this projection, the non-locality 
effect would not be significant, at least for the argument of confinement.)

As another way to clarify the confinement on the periodic lattice, 
we also investigate the Polyakov loop $\langle L_P \rangle \equiv 
\langle {\rm tr} \prod_{t=1}^{L}U_4(\vec x,t)\rangle/3$ 
and the center $Z_3$-symmetry \cite{R05} in terms of the Dirac-mode projection.
The Polyakov loop $\langle L_P \rangle$, which is usually used 
at finite temperature, can be also applied to our temporally periodic system 
on the link-variable, and it physically relates to 
the quark single-particle energy and the $Z_3$-symmetry \cite{R05}.
Note that the non-locality effect is less significant for 
the Polyakov loop $\langle L_P \rangle$ or the quark single-particle energy.

Now, we calculate the Polyakov loop with cutting off 
of the low-lying Dirac modes, 
\begin{eqnarray}
\langle L_P \rangle_{\rm IR} \equiv 
\frac{1}{3}\frac{1}{V}\langle {\rm Tr} (\prod_{k=1}^{L}\hat U_4^P) \rangle
=\frac{1}{3}\frac{1}{V}\langle {\rm Tr} \{(\hat U_4^P)^L\} \rangle,
\end{eqnarray}
and its scatter plot, 
using the same lattice ($6^4, \beta=5.6$) and 
the same IR cutoff $\Lambda_{\rm IR}=0.5a^{-1}$.
In the use of the full Dirac modes, i.e., $\hat P=1$, 
$\langle L_P \rangle_{\rm IR}$ coincides with $\langle L_P \rangle$.
We show in Fig.6 the scatter plot of the Polyakov loop 
$\langle L_P \rangle_{\rm IR}$ after cutting off the low-lying Dirac modes 
below $\Lambda_{\rm IR}=0.5a^{-1}$.
We find that the IR-cut Polyakov loop $\langle L_P \rangle_{\rm IR}$ 
remains to be almost zero, i.e., $\langle L_P \rangle_{\rm IR}\simeq 0$, 
which corresponds to the $Z_3$-unbroken phase. 
In fact, even after removing the low-lying Dirac modes, 
which are responsible to chiral symmetry breaking, 
the single-quark energy is extremely large and 
the system is in the $Z_3$-unbroken confinement phase.

\begin{figure}[h]
\begin{center}
\includegraphics[scale=0.44]{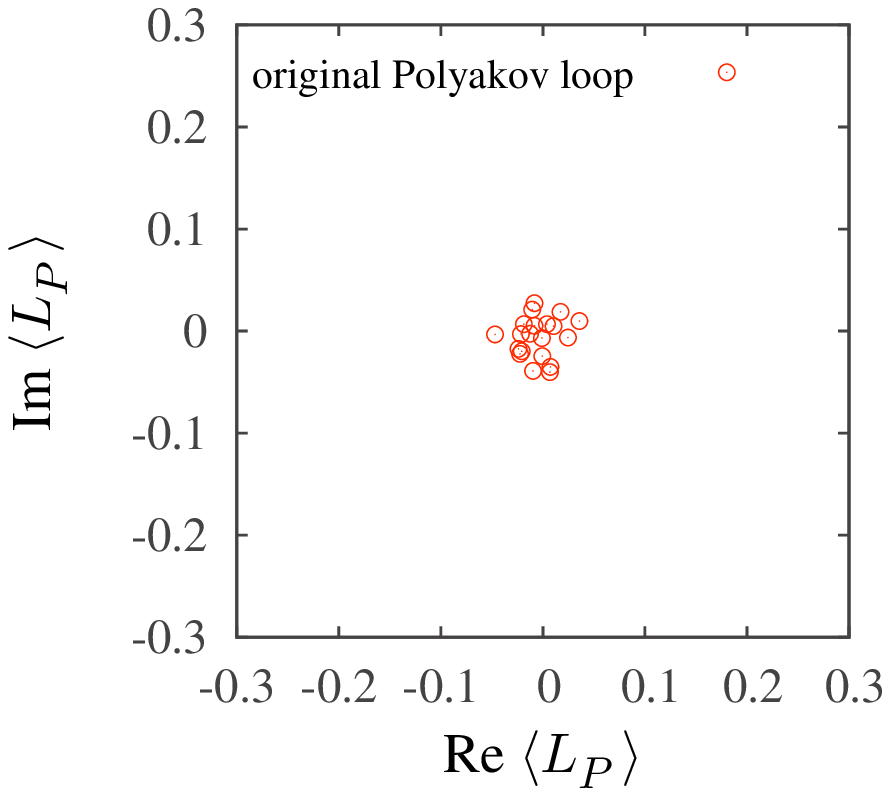}
\includegraphics[scale=0.44]{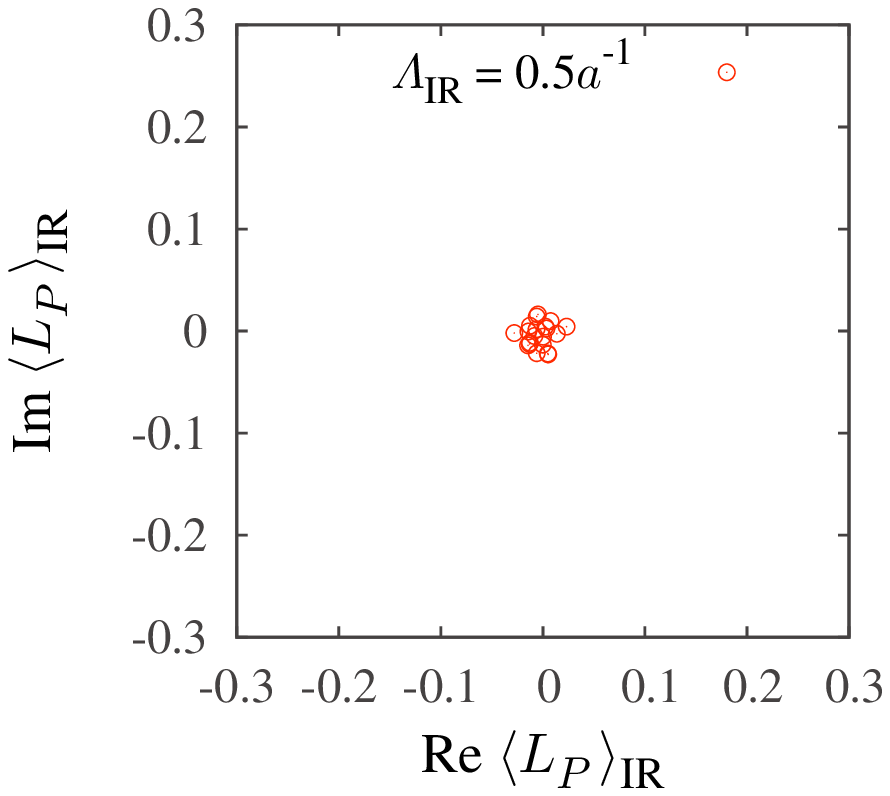}
\caption{
The scatter plot of the Polyakov loop.
The left figure shows the original Polyakov loop $\langle L_P \rangle$.
The right figure shows the Polyakov loop 
$\langle L_P \rangle_{\rm IR}$ after cutting off the low-lying Dirac modes 
below the IR-cutoff $\Lambda_{\rm IR}=0.5a^{-1}$. 
}
\end{center}
\end{figure}

We also investigate the UV-cut of Dirac modes in lattice QCD, 
using the UV-cut Dirac spectral density 
$\rho_{\rm UV}(\lambda)\equiv 
\rho(\lambda)\theta(\Lambda_{\rm UV}-|\lambda|)$ 
with the UV-cutoff $\Lambda_{\rm UV}=2a^{-1}\simeq 1.6{\rm GeV}$.
In this case, unlike the IR cut, 
the chiral condensate is almost unchanged, 
and chiral symmetry breaking is almost kept. 
We show in Fig.7 the UV-cut Wilson loop and 
the corresponding inter-quark potential, 
after removing the UV Dirac modes. 
We find that the area-law behavior of the Wilson loop 
and the slope parameter $\sigma^P$ 
are almost unchanged by the UV-cut of the Dirac modes. 
This result seems consistent with the pioneering lattice study 
of Synatschke-Wipf-Langfeld \cite{SWL08}: 
they found that the confinement potential is almost reproduced 
only with low-lying Dirac modes, 
using the spectral sum of the Polyakov loop \cite{BBGH08,G06BGH07}.

\begin{figure*}[ht]
\begin{center}
\includegraphics[scale=0.4]{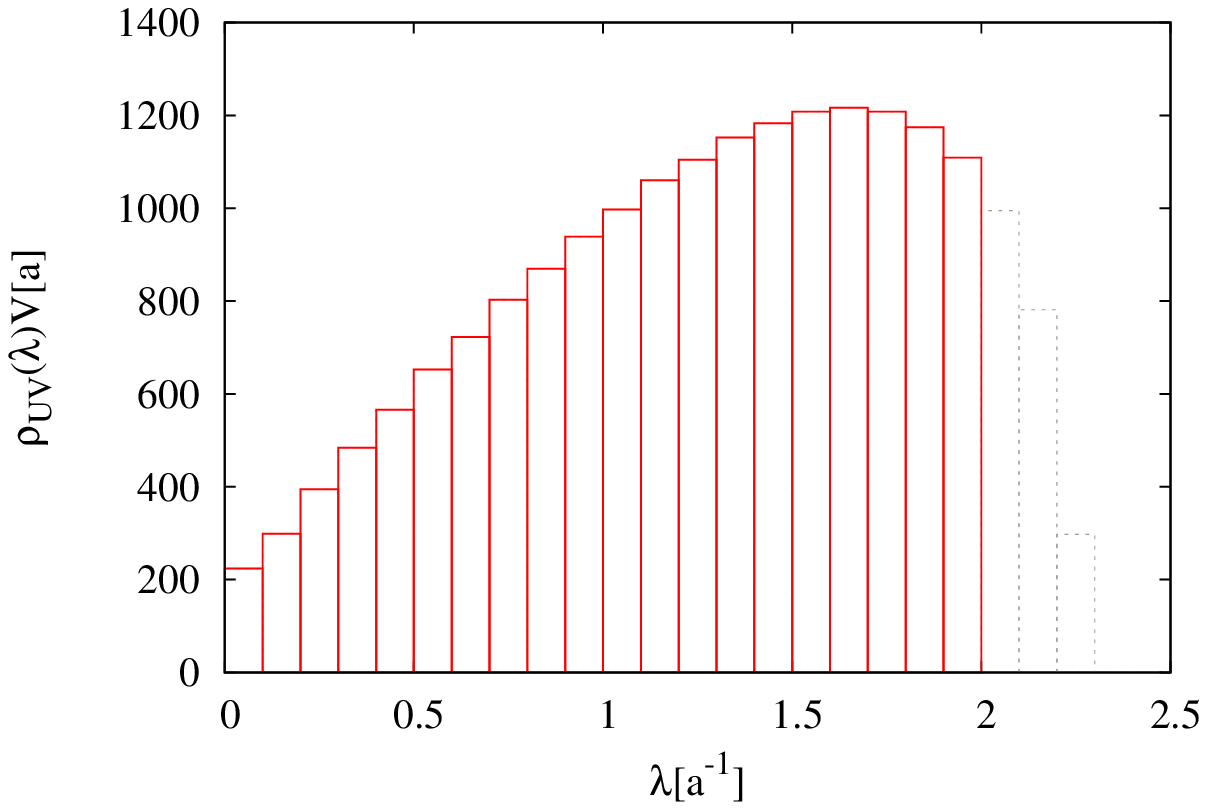}
\includegraphics[scale=0.4]{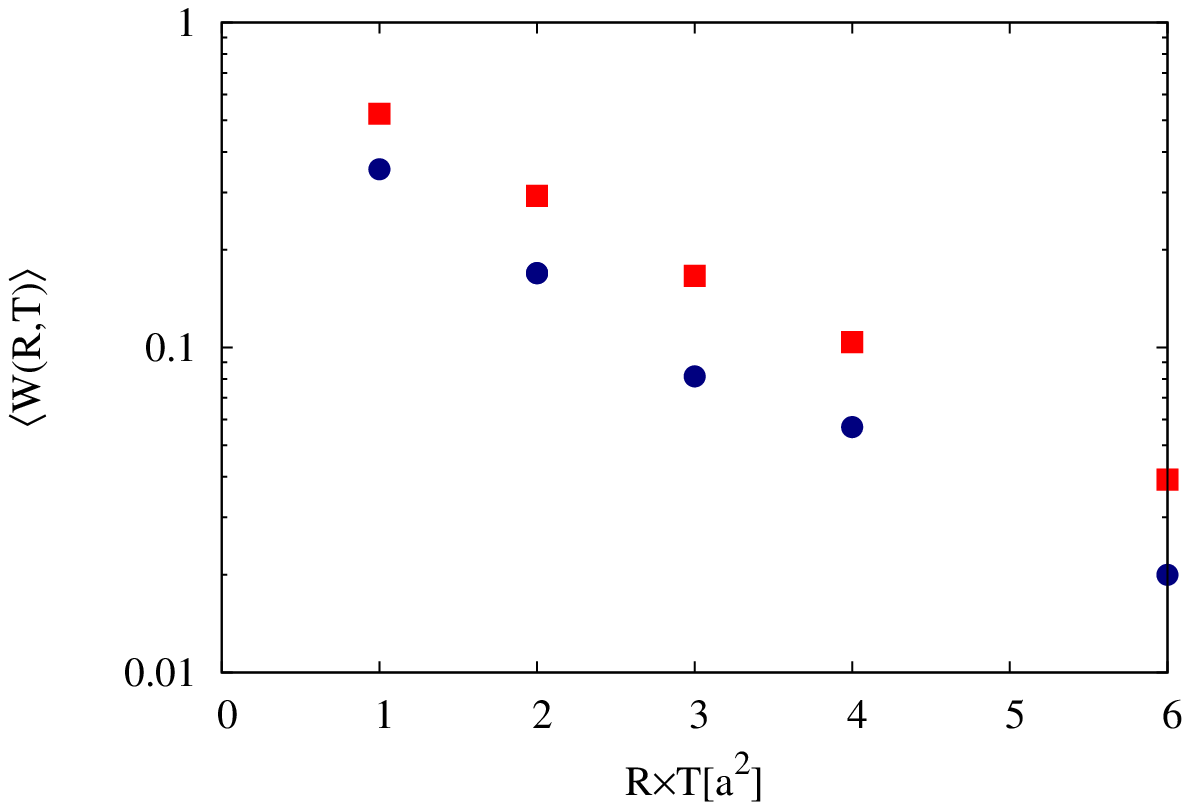}
\includegraphics[scale=0.4]{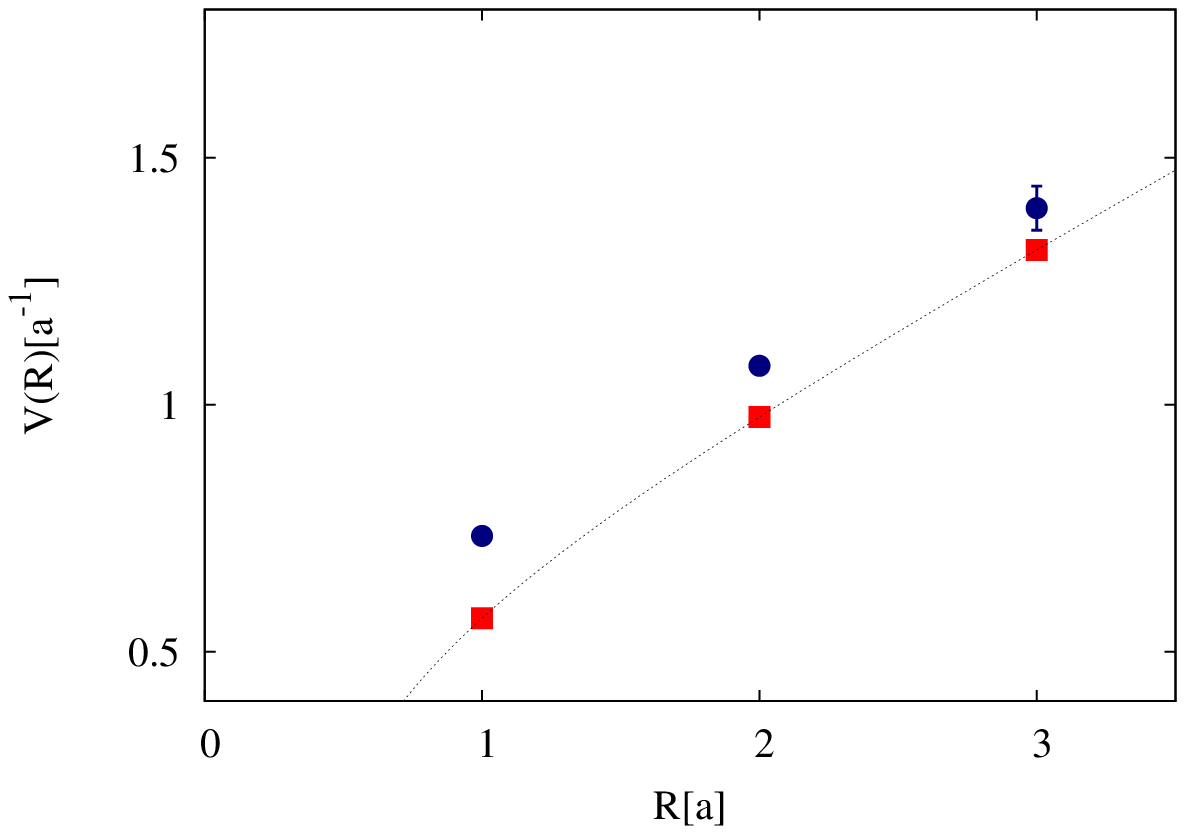}
\caption{
(a) The UV-cut Dirac spectral density 
$\rho_{\rm UV}(\lambda)\equiv 
\rho(\lambda)\theta(\Lambda_{\rm UV}-|\lambda|)$ 
with the UV-cutoff $\Lambda_{\rm UV}=2a^{-1}\simeq 1.6{\rm GeV}$.
(b) The UV-cut Wilson loop ${\rm Tr} W^P(R,T)$ (circle) 
after removing the UV Dirac modes, plotted against $R \times T$. 
The slope parameter $\sigma^P$ is almost the same as that of 
the original Wilson loop (square).
(c) The corresponding UV-cut inter-quark potential (circle), 
which is almost unchanged from the original one (square),
apart from an irrelevant constant.
}
\end{center}
\end{figure*}

Furthermore, we examine ``intermediate(IM)-cut'', where 
a certain part of $\Lambda_1 < |\lambda_n| < \Lambda_2$ 
of Dirac modes is removed. 
Unfortunately, when the wide region of Dirac modes is removed, 
the statistical error becomes quite large 
for the Dirac-mode projected Wilson loop.
Here, we remove the IM Dirac modes of 
$0.5-0.8 [a^{-1}]$, $0.8-1.0 [a^{-1}]$, 
and $1.0-1.2[a^{-1}]$, respectively, 
and investigate the corresponding IM-cut Wilson loop and 
the corresponding inter-quark potential 
in each case, as shown in Fig.8.
For each case, the area-law behavior of the Wilson loop 
and the slope parameter $\sigma^P$ 
are found to be almost unchanged by the IM-cut of the Dirac modes. 

\begin{figure*}[h]
\begin{center}
\includegraphics[scale=0.4]{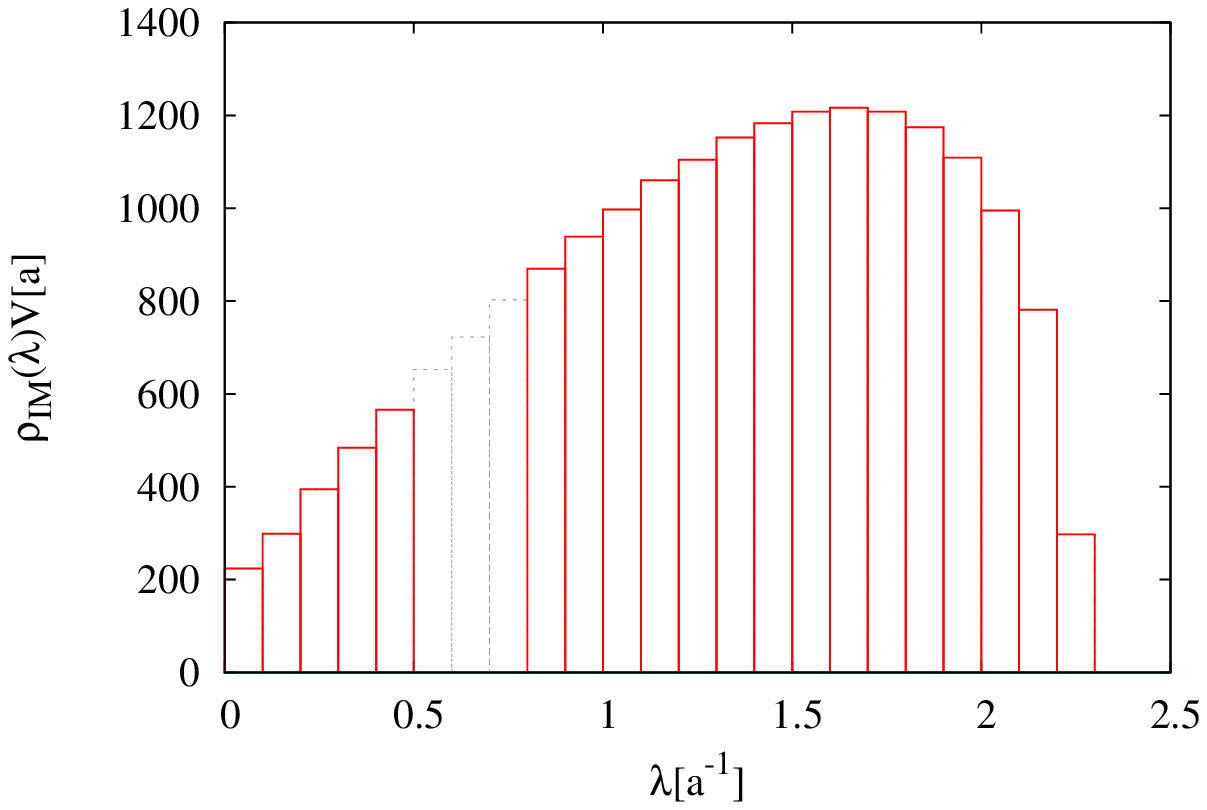}
\includegraphics[scale=0.4]{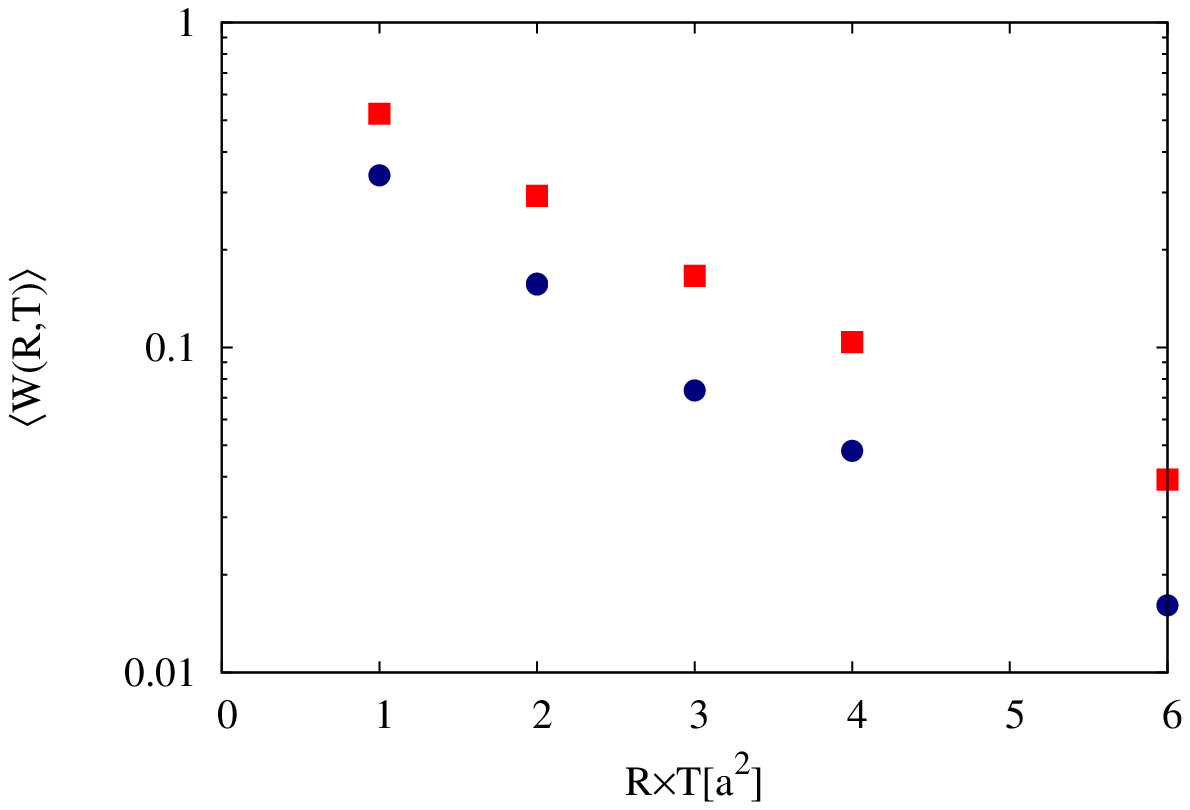}
\includegraphics[scale=0.4]{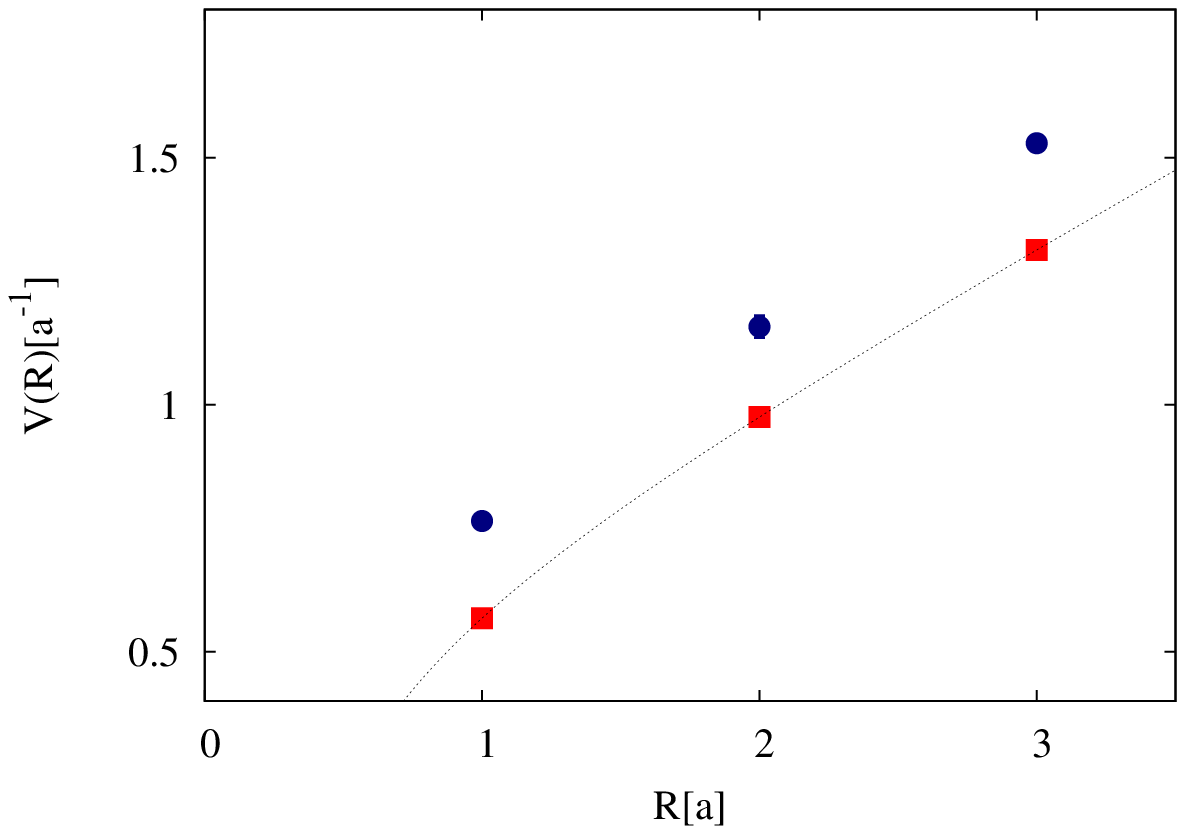}
\includegraphics[scale=0.4]{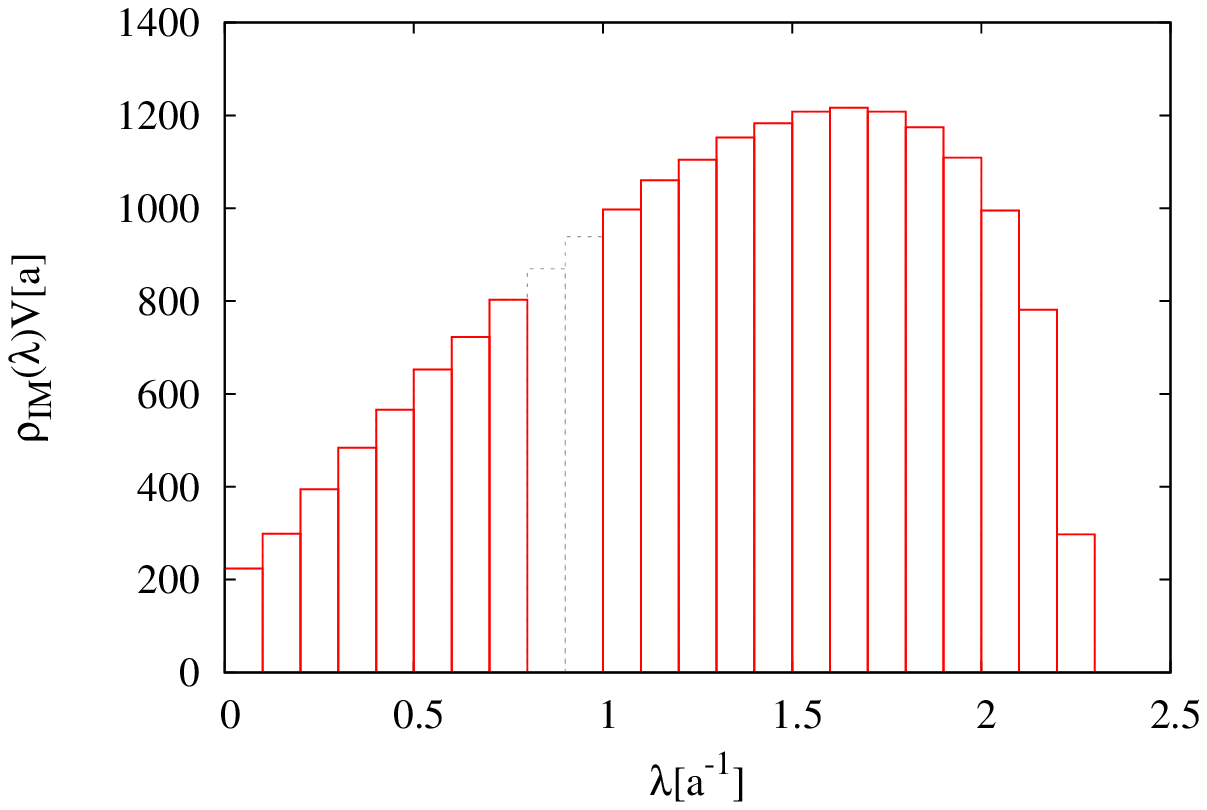}
\includegraphics[scale=0.4]{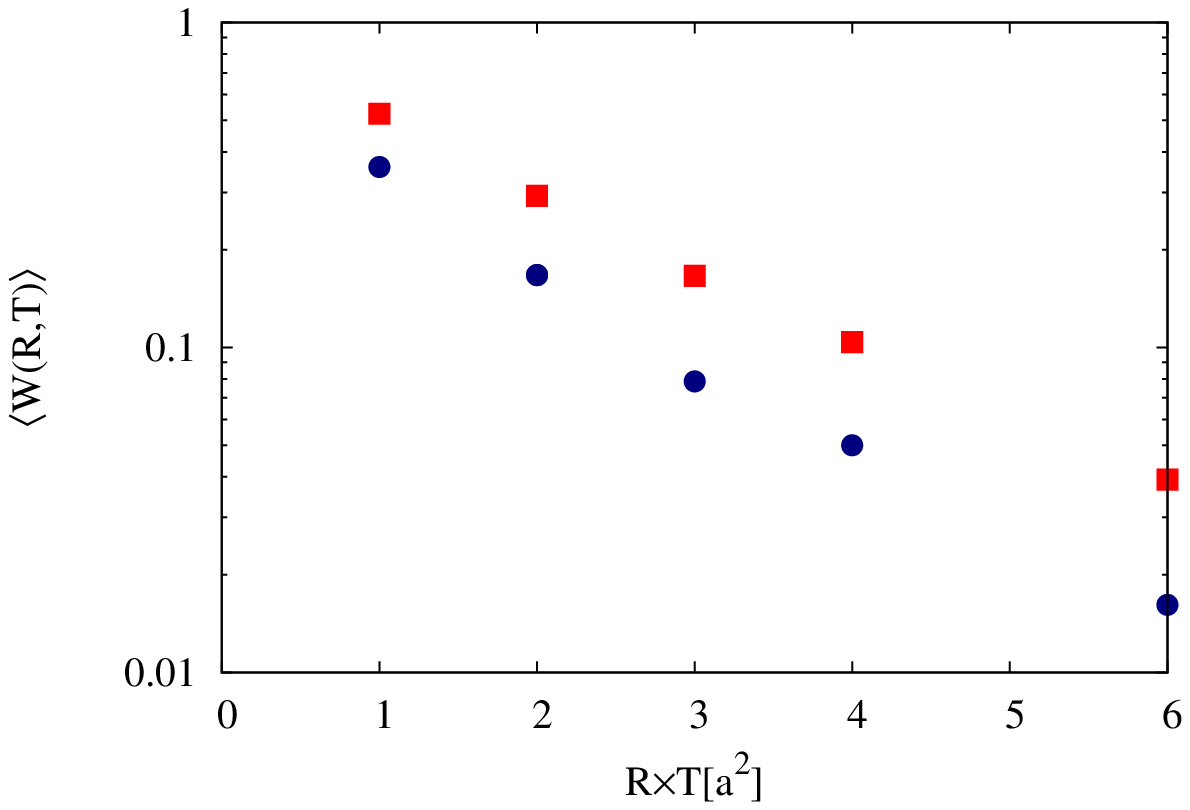}
\includegraphics[scale=0.4]{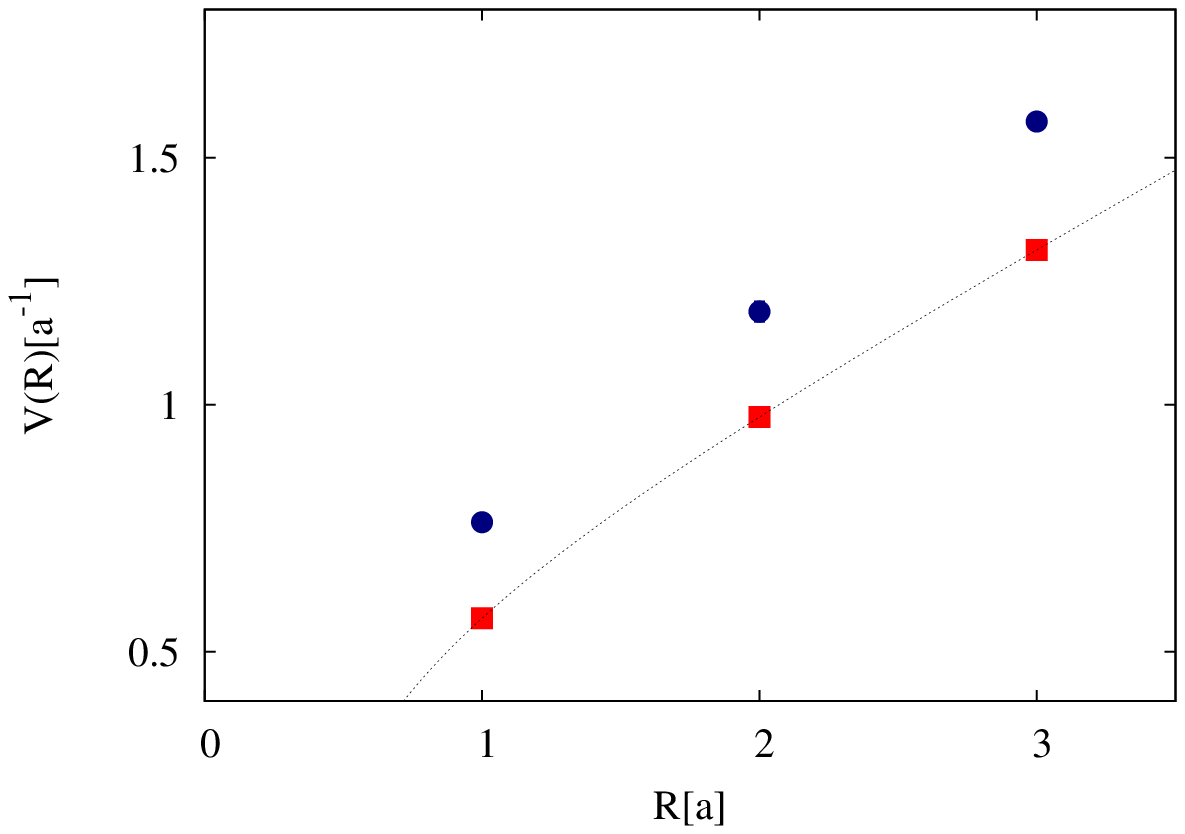}
\includegraphics[scale=0.4]{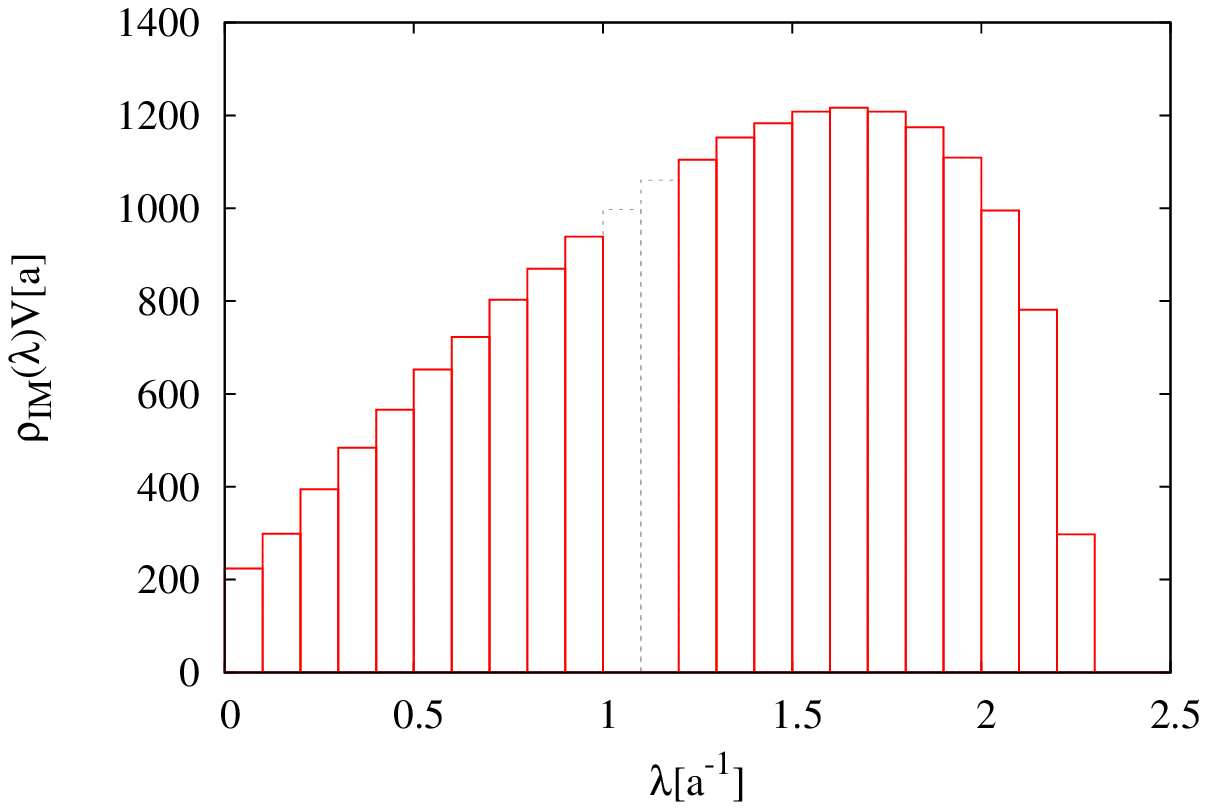}
\includegraphics[scale=0.4]{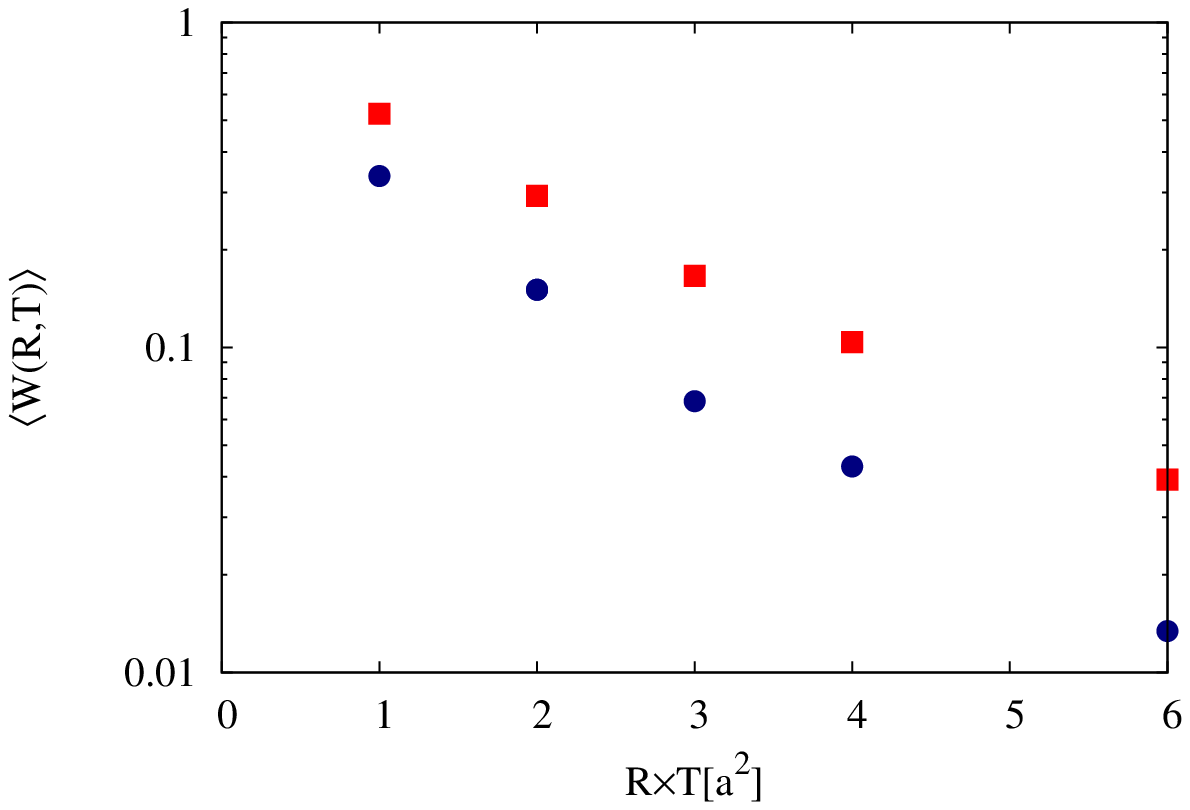}
\includegraphics[scale=0.4]{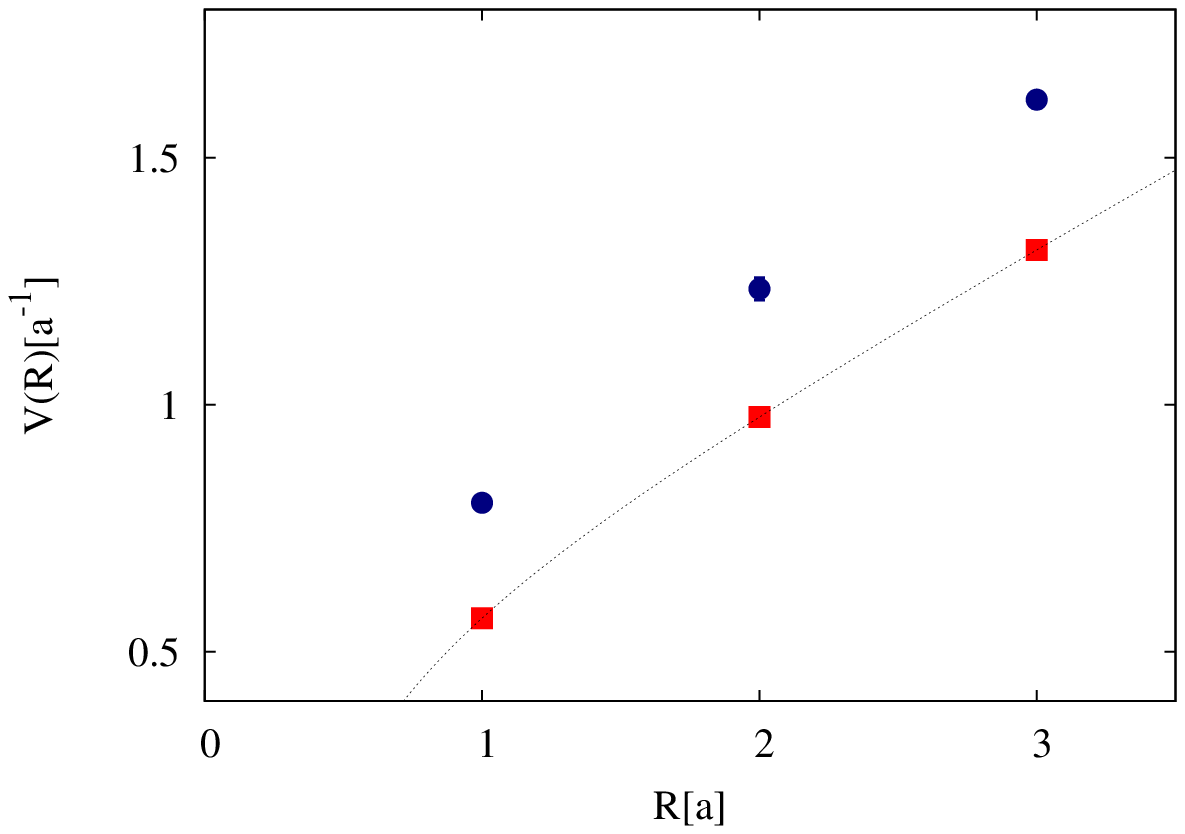}
\caption{
The left figures show 
the intermediate(IM)-cut Dirac spectral density $\rho_{\rm IM}(\lambda)$:
the IM Dirac modes of 
$0.5-0.8 [a^{-1}]$ (top), $0.8-1.0 [a^{-1}]$ (middle), and 
$1.0-1.2[a^{-1}]$ (bottom) are cut. 
The central figures show the 
IM-cut Wilson loop ${\rm Tr} W^P(R,T)$ (circle) 
after removing the IM Dirac modes, 
plotted against $R \times T$. 
For each case, the slope parameter $\sigma^P$ is 
almost the same as that of the original Wilson loop (square).
The right figure shows the corresponding 
IM-cut inter-quark potential (circle), which is almost 
unchanged from the original one (square),
apart from an irrelevant constant.
}
\end{center}
\end{figure*}

Thus, from the above lattice QCD results, 
we conclude that there is no specific region of 
the Dirac modes responsible to the confinement in QCD, 
unlike the chiral symmetry breaking.
Instead, we conjecture that the ``seed'' of confinement is 
distributed not only in low-lying Dirac modes but also 
in a wider region of the Dirac-mode space. 

\section{Summary and Discussions}

We have developed a manifestly gauge-covariant expansion and projection 
using the eigen-mode of the QCD Dirac operator 
$\Slash D=\gamma^\mu D^\mu$. 
With this method, we have performed a direct investigation 
of correspondence between confinement and chiral symmetry breaking 
in SU(3) lattice QCD on the $6^4$ periodic lattice 
at $\beta$=5.6 at the quenched level. 
We have found that 
the Wilson loop remains to obey the area law, 
and the slope parameter corresponding to the string tension or 
the confinement force is almost unchanged, 
even after removing the low-lying Dirac modes, 
which are responsible to chiral symmetry breaking.
We have also found that the Polyakov loop remains to be almost zero 
even without the low-lying Dirac modes, 
which indicates the $Z_3$-unbroken confinement phase.
These results indicate that one-to-one correspondence does not 
hold for between confinement and chiral symmetry breaking in QCD.

As a caution, we have used a coarse and small lattice, 
because of the technical difficulty to diagonalize the full Dirac operator.
In particular, the box size of our lattice volume is about 1.5fm.
In fact, to be strict, this region we survey is the intermediate distance, 
of which confining behavior is rather important for the quark-hadron physics.
To obtain more definite conclusion, especially 
on the asymptotic confining behavior of the potential, 
it is desired to perform larger-volume lattice QCD calculations 
and to cut various wider region of the Dirac modes, 
although it is technically quite difficult.

Our strategy is to investigate the relation 
between the nonperturbative properties of QCD, 
by extracting or removing 
the essence of chiral symmetry breaking.
This is similar to the demonstration of Abelian/monopole dominance 
\cite{tH81,SST95,M95,W95,KSW87,SNW94,STSM95,KMS95,AS99,EI82}
or center/vortex dominance \cite{HFGHO08,DFGO97,KT98} 
for nonperturbative properties.
However, while the previous scenario has been done 
in a specific gauge, our new method is manifestly gauge-invariant.
In this analysis, we have carefully amputated only the 
``essence of chiral symmetry breaking'' 
by cutting off the low-lying Dirac modes.
Then, we have artificially realized the 
``confined but chiral restored situation'' in QCD.

Recently, Lang and Schrock studied the hadron spectra 
after the cut of the low-lying Dirac modes \cite{LS11}.
Since the quark propagator is directly expressed 
with the Dirac operator $\Slash D$, 
the Dirac-mode projection is straightforward, 
and complicated projection procedure is not necessary in such studies.
In their study, although the confinement was not checked, 
the appearance of hadronic spectra seems to suggest 
the existence of the confinement force, 
even after cutting the low-lying Dirac modes.

Next, we comment on the possible relation among 
confinement, chiral symmetry breaking, and monopoles in QCD. 
There is a close relation 
between confinement and chiral symmetry breaking 
through the monopoles in the MA gauge \cite{SST95,M95,W95}.
The monopole would be essential degrees of freedom for 
most nonperturbative QCD: confinement \cite{SNW94}, 
chiral symmetry breaking \cite{M95,W95}, and instantons \cite{STSM95}.
In fact, removing the monopole would be ``too fatal''
for the nonperturbative properties, so that 
nonperturbative QCD phenomena are simultaneously lost by their cut. 
On the approximate coincidence of 
the critical temperatures of deconfinement and chiral restoration, 
a large change of monopoles may lead to both phase transitions \cite{M95},
since the global connection of the monopole current 
seems to be largely changed around the QCD phase transition \cite{KMS95}. 

As for the recent finite-temperature QCD analysis, 
a lattice QCD group has reported 
a certain difference between the ``critical temperatures'' of 
deconfinement and chiral restoration, 
which are determined by the susceptibility peak of 
the Polyakov loop and chiral condensate, respectively \cite{AFKS06}.
This may be also an indirect evidence of 
``confinement $\ne$ chiral symmetry breaking'' in QCD.

Next, we briefly discuss the role of low-lying Dirac modes 
in the viewpoint of instantons in QCD.
The Dirac zero-mode associated with an instanton 
is localized around it \cite{S94}. 
However, the localized objects are hard to contribute to 
the large-distance phenomenon of confinement, 
although such low-lying Dirac modes contribute to 
chiral symmetry breaking.
Recall that instantons contribute to chiral symmetry breaking,
but do not directly lead to confinement \cite{S94}. 
Then, as a thought experiment, 
if only instantons can be carefully removed from the QCD vacuum, 
confinement properties would be almost unchanged, 
but the chiral condensate is largely reduced, 
and accordingly some low-lying Dirac modes disappear. 
Thus, in this case, confinement is almost unchanged, 
in spite of the large reduction of low-lying Dirac modes.

If the relation between confinement and chiral symmetry breaking 
is not one-to-one in QCD, richer phase structure is expected in QCD. 
For example, the phase transition point can be different 
between deconfinement and chiral restoration 
in the presence of strong electro-magnetic fields, 
because of their nontrivial effect on chiral symmetry \cite{ST9193}.
It is also interesting to investigate the similar analysis 
at finite temperatures in lattice QCD.
The full QCD calculation in this direction is also an interesting subject.

\section*{Acknowledgements}
The authors thank Prof. K.~Langfeld and Dr. T.~Misumi for useful comments.
H.S. is supported in part by the Grant for Scientific Research 
[(C) No.~23540306, Priority Areas ``New Hadrons'' (E01:21105006)] 
and T.I. is supported by Grant-in-Aid for JSPS Fellows (No.23-752), 
from the Ministry of Education, Culture, Science and Technology 
(MEXT) of Japan. 
This work is supported by the Global COE Program, 
``The Next Generation of Physics, Spun from Universality and Emergence".
The lattice QCD calculations are done on NEC SX-8R at Osaka University.

\end{document}